\documentclass{bmcart}

\usepackage{amsthm,amsmath}
\usepackage[utf8]{inputenc} 
\usepackage{caption}
\usepackage{braket}
\usepackage{multicol}
\usepackage{subcaption}
\usepackage{graphicx}

\startlocaldefs
\endlocaldefs

\begin{document}

\begin{frontmatter}

\begin{fmbox}
\dochead{Research}


\title{ResQNets: A Residual Approach for Mitigating Barren Plateaus in Quantum Neural Networks}


\author[
  addressref={aff1},                   
  corref={aff1},                       
  email={mkashif@hbku.edu.qa}   
]{\inits{M.K.}\fnm{Muhammad} \snm{Kashif}}
\author[
  addressref={aff1},
  email={smalkuwari@hbku.edu.qa}
]{\inits{S.A.K.}\fnm{Saif} \snm{Al-Kuwari}}


\address[id=aff1]{
  \orgdiv{Division of Information and Computing Technology, College of Science and Engineering},             
  \orgname{ Hamad Bin Khalifa University, Qatar Foundation},          
  \city{Doha},                              
  \cny{Qatar}                                    
}



\end{fmbox}


\begin{abstractbox}

\begin{abstract} 
The barren plateau problem in quantum neural networks (QNNs) is a significant challenge that hinders the practical success of QNNs. In this paper, we introduce residual quantum neural networks (ResQNets) as a solution to address this problem. ResQNets are inspired by classical residual neural networks and involve splitting the conventional QNN architecture into multiple quantum nodes, each containing its own parameterized quantum circuit, and introducing residual connections between these nodes. Our study demonstrates the efficacy of ResQNets by comparing their performance with that of conventional QNNs and plain quantum neural networks through multiple training experiments and analyzing the cost function landscapes. Our results show that the incorporation of residual connections results in improved training performance. Therefore, we conclude that ResQNets offer a promising solution to overcome the barren plateau problem in QNNs and provide a potential direction for future research in the field of quantum machine learning.
\end{abstract}




\end{abstractbox}
%

\end{frontmatter}



\section{Introduction}

The Noisy Intermediate-Scale Quantum (NISQ) devices are a new generation of quantum computers capable of executing quantum algorithms. However, NISQ devices still suffer from significant errors and limitations in terms of the number of qubits and coherence time \cite{Bharti:2022}. 
Despite these limitations, NISQ devices are an important stepping stone towards developing fault-tolerant quantum computers, as they provide a platform for exploring and evaluating basic quantum algorithms and applications \cite{Preskill:2018,schuld:2019}. 
Research in the NISQ era is focused on developing algorithms and techniques that are resilient to noise and errors, and can run effectively on NISQ devices \cite{Lau:2022}. This includes algorithms for quantum error correction \cite{Roffe:2019}, quantum optimization \cite{Moll:2018}, and quantum machine learning (QML)\cite{Biamonte:2017}.

QML is an interdisciplinary field that combines the concepts and techniques from quantum computing and machine learning (ML).
It aims to leverage the unique properties of quantum systems, such as superposition, entanglement, and interference, to develop new algorithms and approaches for solving complex machine learning problems \cite{Schuld:2014,Mitarai:2018}.
QML is increasingly becoming an exciting application in the NISQ era \cite{Preskill:2018}. 
The anticipation here is that the quantum models (by exploiting the exponentially large Hilbert space) would achieve a computational advantage over their classical counterparts \cite{Liu:2021,Huang:2022}, particularly for quantum datasets \cite{Cong:2019,Schatzki:2021,Caro:2022}. 
With continued advancements in quantum hardware \cite{Nathalie:2021}, development of new quantum algorithms \cite{Lloyd:2014}, quantum error correction and fault tolerance \cite{Norbert:2017}, the future of QML is bright, and it is likely to play a significantly important role in the field of machine learning. 
A wide range of ML algorithms are being explored in the quantum realm, including quantum neural networks (QNNs) \cite{Abel:2022}, quantum support vector machines \cite{Rebentrost:2014,Havlicek:2019}, quantum principal component analysis \cite{Lloyd:2014a}, and quantum reinforcement learning \cite{Dunjko:2017,meyer:2022,lockwood:2020}. 
These approaches were shown to be effective in various domains, such as image classification \cite{Farhi:2018,mari:2020,mathur:2021,Cong:2019,Pesah:2021,Chen:2022}, natural language processing \cite{Meichanetzidis:2021,coecke:2020,Sipio:2022}, and recommendation systems \cite{Gao:2023}.

QNNs is a promising area of research that aims to combine the power of quantum computing and neural networks to solve complex computational problems \cite{Wan:2017,Killoran:2019}. 
Unlike classical neural networks, QNNs use quantum-inspired representations and operations for encoding and processing data \cite{Zoufal:2019,beer:2020,Kashif:2021}. 
This allows for the exploration of exponential solution space and the exploitation of quantum parallelism, potentially leading to faster and more accurate results \cite{Schuld:2014,Yuxuan:2020,Caro:2022,Kashif:2022capcity}. 
QNNs can be considered as a subclass of variational quantum algorithms, which aim to optimize parameters $(\theta)$ of a parameterized quantum circuit (PQC) \footnote{we will use the terms ``PQC" and ``quantum layers" interchangeably} $U(\theta)$ to minimize the cost function $\mathcal{C}$. PQC utilizes tunable parameters to optimize quantum algorithms through classical computation.
One example of a QNN architecture is the quantum Boltzmann machine \cite{Amin:2018,Zoufal:2021}, which uses quantum circuits to model complex probability distributions and perform unsupervised learning tasks.
In addition to unsupervised learning, QNNs have shown potential in various applications such as quantum feature detection \cite{Havlicek:2019}, quantum data compression and denoising\cite{Romero:2017,Bondarenko:2020}, and quantum reinforcement learning \cite{Kwak:2021,chen:2020}. QNNs can also be used for quantum-enhanced image recognition \cite{Biamonte:2017,Banchi:2020} and quantum molecular simulations \cite{Grimsley:2019}.

However, despite their potential, QNNs are still in the early stages of development and face several technical and practical challenges.  
In particular, training and optimizing the parameters in QNNs pose significant challenges. To address these challenges, the research community has been developing the quantum landscape theory \cite{Arrasmith:2022} that explores the properties of cost function landscapes in QML systems. 
Consequently, interesting results have been obtained in the study of QNN's training landscapes, including the occurrence of barren plateaus (BP) \cite{McClean:2018}, the presence of sub-optimal local minima \cite{Wierichs:2020}, and the impact of noise on cost function landscapes \cite{Wang:2021,Fontana:2020,wang:2021c,Franca:2021}. These findings provide important insights into the properties of QNNs and their training dynamics, and can inform the development of new algorithms and strategies for training and optimizing QNNs.

In particular, the BP problem refers to a phenomenon in which the circuit's expressiveness, as measured by its ability to approximate a target unitary operation, is severely limited as the number of qubits in the circuit increases \cite{McClean:2018}, which is mainly due to vanishing gradients in the parameter space. 
The phenomenon of BP in QNNs is a significant challenge that impedes the advancement and widespread implementation of QNNs. 
To mitigate the BP, various strategies have been proposed, including the use of clever parameter initialization techniques \cite{Liu:2023}, pre-training \cite{Verdon:2019}, examination of the dependence on the cost function \cite{Cerezo:2021aa,Kashif:2023}, implementation of layer-wise training of QNNs \cite{Skolik:2021}, and initialization parameters drawn from the beta distribution\cite{Kulshrestha:2022}. The trainability vs expressibility analysis of QNNs from the aspect of BP is conducted in \cite{kashif:2023unified}, where a trade-off between quantum layers width and depth has been observed for a better learning performance.   
These solutions aim to overcome the limitations posed by the BP in QNNs and facilitate the full realization of their potential. However, it is important to note that the solution that works best for one QNN architecture may not work for another, as the BP problem can be highly dependent on the specific problem being solved and the quantum architecture being used.

\paragraph{Related Work.}
In recent research efforts, the concept of utilizing the residual approach in QNNs has gained traction. One such work, proposed in \cite{LIANG:2021}, introduces a hybrid quantum-classical neural network with deep residual learning. The authors explore the integration of the residual block structure into QNN architecture and highlight its potential benefits. Specifically, they emphasize that connecting residual blocks with QNNs can enhance robustness against noise, a crucial consideration in quantum computing applications.

In contrast, our research focuses on a different aspect of the residual approach in QNNs. We aim to transform the traditional QNN architecture into a residual structure by dividing the conventional QNN architecture into multiple quantum nodes (each containing its own PQC), and investigate its effectiveness in mitigating the BP phenomenon. BP are a challenge that arises as the number of qubits in quantum circuit increases, leading to deep quantum circuits. Our primary objective is to address this issue and improve the training performance of QNNs.

Furthermore, a related study in \cite{ABDELAZIZ:2022} employs the residual approach in the optimization of an IoT platform. The authors also conclude that the residual approach in QNNs exhibits greater robustness against noisy data and better performance in learning unitary functions.

In a different context, \cite{ResQNet_thesis} explores the residual approach in QNNs but with a focus on shallower quantum circuits rather than deep ones. The authors aim to achieve comparable performance with shallower circuits, and they suggest manipulating data encoding strategies to improve accuracy. Our work, however, entirely concentrates on quantum circuit width, i.e., the number of qubits, as a means to study and address the barren plateaus phenomenon, independent of the data encoding technique.

Overall, these research efforts collectively contribute to the growing body of knowledge on the residual approach in QNNs, highlighting its potential benefits for various quantum computing applications.

\paragraph{Contribution.}
In this paper, we propose a novel solution for mitigating the issue of barren plateaus (BP) in quantum neural networks (QNNs). 
Our methodology draws inspiration from the framework of residual neural networks, which were proposed to address the issue of vanishing gradients in classical neural networks.
In this context, we introduce the concept of residual quantum neural networks (ResQNets) by incorporating residual connections between two quantum layers of variable depths. Our findings suggest that the utilization of ResQNets substantially enhances the training process of QNNs as compared to their non-residual counterparts, denoted as PlainQNets. 
To substantiate the efficacy of our proposed ResQNets, we undertake a systematic comparison involving an analysis of the cost function landscapes and an assessment of their training performance with that of PlainQNets. 
The results obtained from our experimental investigations elucidate that the incorporation of residual connections in QNNs (ResQNets) effectively mitigate the adverse effects of BP and result in improved overall training performance.

\paragraph{Organization.}
The rest of the paper is organized as follows: Section \ref{sec:ResidualNNs} provides an overview of both classical and quantum residual neural networks and motivates their application.  Section \ref{sec:PQCs} discusses parameterized quantum circuits and elaborate on how can multiple PQCs be cascaded. This section also introduces the residual approach in cascaded PQCs. 
The methodology we adopt in this paper while conducting the various experiments is provided in Section \ref{sec:methodology}. Section \ref{sec:results} presents the results we obtained on both the simulation environment and real quantum devices. Finally, the paper concludes in section \ref{sec:conclusion} with a few concluding remarks and pointers to possible extensions to this work. 


\section{Residual Neural Networks}\label{sec:ResidualNNs}

Residual Neural Networks (ResNets) are a type of deep neural network architecture that aims to improve the training process by addressing the vanishing gradient problem. The basic idea behind ResNets is to introduce residual connections between layers in the network, allowing for easier optimization as the network gets deeper. The residual connections allow the network to learn residual mapping rather than trying to fit the target function directly. This helps prevent the vanishing gradient problem, where the gradients in the backpropagation process become very small, making it difficult to update the parameters effectively.
ResNets were first introduced in \cite{Kaiming:2016}, where the authors showed that ResNets outperformed traditional deep neural networks on benchmark image recognition tasks and demonstrated that ResNets could accommodate significantly deeper architectures than previous networks without sacrificing accuracy.

\begin{figure*}[h!]
     
        \begin{subfigure}[b]{0.75\textwidth}
         \centering
        \includegraphics[scale=0.5]{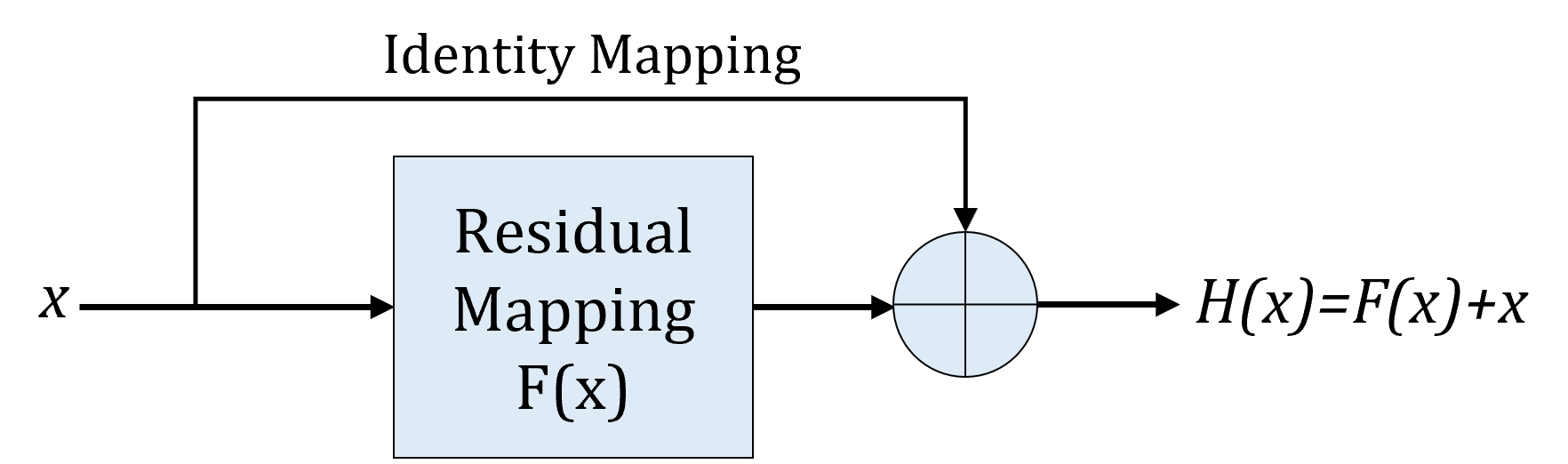}
         \caption{ResNets}
         \label{fig:residual_structure_CNN}
     \end{subfigure}
       \hfil\quad
   \begin{subfigure}[b]{0.75\textwidth}
         \centering
         \hspace{0.5cm}
         \includegraphics[scale=0.5]{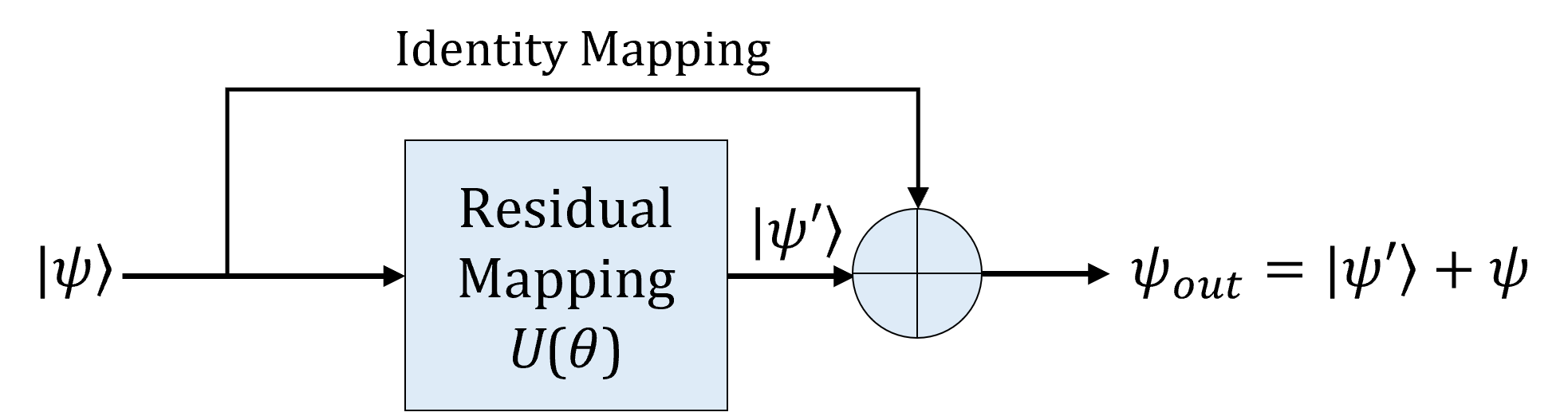}
         \caption{ResQNets}
        \label{fig:ReQNet_structure}
     \end{subfigure}
     \captionsetup{hypcap=false}
    \caption{Residual block structure}  
    \label{fig:ResNet_ResQNet_strcuture}
    
\end{figure*}

The residual connections in ResNets have been shown to be effective for training very deep neural networks, with hundreds or even thousands of layers. This has drastically improved the performance in several computer vision and natural language processing tasks.
A typical structure of a residual block is depicted in Fig. \ref{fig:residual_structure_CNN}.

Given an input feature map $x$, the basic building block of a ResNet can be defined as:

$$H(x) = F(x, {W_i}) + x$$

\noindent where $H(x)$ is the output of the block, $F$ is a non-linear function represented by a series of neuron and activation layers with parameters ${W_i}$, and $x$ is the input feature map that is added back to the output (the residual connection). The model is trained to learn the function $F$ such that it approximates the residual mapping $y - x$, where $y$ is the desired output.
By introducing residual connections, ResNets can address the vanishing gradient problem in deep neural networks, allowing for deeper architectures to be trained effectively.

In this paper, we introduce the quantum counterpart of ResQNet, namely residual quantum neural network (ResQNet), a QNN architecture combining the principles of classical ResNets with QNNs. The basic idea is to add a residual connection between the output of one layer of quantum operations and the input of the next layer. This helps to mitigate the vanishing gradient problem, a.k.a BP, which is a major challenge in QNNs and arises as the number of qubits in the systems increases. Fig. \ref{fig:ReQNet_structure} directs how ResQNets is compared to ResNets. 

In ResQNets, the residual connection is represented mathematically as:

$$ \psi_\mathrm{out}(\theta) = \psi(\theta) + U(\theta) \psi(\theta) $$

\noindent where $\psi(\theta)$ is the input to the quantum circuit, $U(\theta)$ is the unitary operation defined by the PQC, and $\psi_\mathrm{out}(\theta)$ is the output.

\section{Parameterized Quantum Circuits} \label{sec:PQCs}
QNN is a type of Parameterized Quantum Circuit (PQC), which is a quantum circuit that has tunable parameters that can be optimized to perform specific tasks. 
In a QNN, the parameters are typically optimized using classical optimization algorithms to learn a target function or perform a specific task. 
The PQC architecture of a QNN allows for the representation and manipulation of quantum data in a manner that can be used for various applications, such as QML and quantum control.
The mathematical derivation of PQC involves the representation of quantum states and gates as matrices and the composition of these matrices to form the overall unitary operator for the circuit.

A quantum state can be represented by a column vector in a Hilbert space, where the elements of the vector are complex numbers that satisfy the normalization constraint:

$$\left\lvert\psi\right\rangle = \begin{bmatrix} \alpha \ \beta \end{bmatrix}, \quad \left\lvert\alpha\right\rvert^2 + \left\lvert\beta\right\rvert^2 = 1$$

A quantum gate is represented by a unitary matrix, which preserves the norm of the vector, i.e., the inner product of the transformed vector with itself is equal to the inner product of the original vector with itself:

$$U^\dagger U = U U^\dagger = I$$

\noindent where $U^\dagger$ is the conjugate transpose of $U$ and $I$ is the identity matrix.
A PQC can be modeled as a sequence of gates, each represented by a unitary matrix based on classical parameters. The overall unitary operator of the circuit can be obtained by composing the matrices of the individual gates in the correct order:

$$U_{\text{circuit}} = U_n(\theta_n)\cdots U_2(\theta_2)U_1(\theta_1)$$

\noindent where $U_i(\theta_i)$ is the unitary matrix representing the $i$-th gate and $\theta_i$ is a classical parameter.

The final quantum state after applying the PQC to an initial state can be obtained by matrix-vector multiplication:

$$\left\lvert\psi_{\text{final}}\right\rangle = U_{\text{circuit}}\left\lvert\psi_{\text{initial}}\right\rangle$$

The parameters $\theta_1,\dots,\theta_n$ can be optimized using classical optimization algorithms to achieve a desired quantum state or to maximize an objective function such as the expected value of a measurement outcome. The optimization problem can be written as:

$$\theta^* = \arg\max_{\theta} \left\lvert\left\langle\psi_{\text{desired}}\right\vert U_{\text{circuit}}(\theta)\left\lvert\psi_{\text{initial}}\right\rangle\right\rvert^2$$

Solving this optimization problem provides the optimal set of parameters $\theta^*$ that produce the desired outcome.


\subsection{Cascading PQCs}
In the proposed ResQNets, we encapsulate PQC/QNNs into a quantum node (QN), and arrange multiple QNs in a series, such that the output from one QN serves as the input for the next. 
This structure enables us to introduce the residual learning approach in a manner that allows the PQCs to work together to achieve the desired outcome.
The process of cascading PQCs involves feeding the output of each PQC into the input of the next, creating a layered structure where each layer represents a single PQC. In this case, each PQC can build upon the outputs of the previous ones, leading to a more complex and sophisticated computation. 
To ensure that the overall computation remains stable, the residual learning approach is employed, where the output of each PQC is combined with the input of the next in a specified manner.

We now present the mathematical formulation for connecting multiple PQCs in sequence. We will refer to each PQC as $U_i$ where $i$ denotes the QN it is encapsulated in. 
%
%

\subsubsection{2-Cascaded PQC}
Consider two PQCs denoted as $U_{1}(\theta_1)$ and $U_{2}(\theta_2)$, where $\theta_1$ and $\theta_2$ are classical parameters. The first PQC $U_{1}(\theta_1)$ is applied to an initial quantum state $\left\lvert\psi_{\text{initial}}\right\rangle$ to obtain an intermediate quantum state $\left\lvert\psi_{\text{intermediate}}\right\rangle$:

$$\left\lvert\psi_{\text{intermediate}}\right\rangle = U_1(\theta_1)\left\lvert\psi_{\text{initial}}\right\rangle$$

The second PQC $U_2(\theta_2)$ is then applied to the intermediate state $\left\lvert\psi_{\text{intermediate}}\right\rangle$ to obtain the final quantum state $\left\lvert\psi_{\text{final}}\right\rangle$:

$$\left\lvert\psi_{\text{final}}\right\rangle = U_2(\theta_2)\left\lvert\psi_{\text{intermediate}}\right\rangle$$

The overall unitary operator of the two cascaded PQCs can be obtained by composing the matrices of the individual PQCs in the correct order:

$$U_{\text{circuit}} = U_2(\theta_2)U_1(\theta_1)$$

The final quantum state after applying the two cascaded PQCs to an initial state can be obtained by matrix-vector multiplication:

$$\left\lvert\psi_{\text{final}}\right\rangle = U_{\text{circuit}}\left\lvert\psi_{\text{initial}}\right\rangle$$

The parameters $\theta_1$ and $\theta_2$ can be optimized using classical optimization algorithms to achieve a desired quantum state or to maximize an objective function such as the expected value of a measurement outcome. The optimization problem can be written as:

    $$\theta_1,\theta_2 = \arg\max_{\theta_1,\theta_2} \left\lvert\left\langle\psi_{\text{desired}}\right\vert U_{\text{circuit}}(\theta_1,\theta_2)\left\lvert\psi_{\text{initial}}\right\rangle\right\rvert^2 = \arg\max_{\theta_1,\theta_2} \left\lvert\left\langle\psi_{\text{desired}}\right\vert U_2(\theta_2)U_1(\theta_1)\left\lvert\psi_{\text{initial}}\right\rangle\right\rvert^2 $$

Solving this optimization problem returns the optimal set of parameters $(\theta_1,\theta_2)$ that produce the desired outcome.


\subsubsection{n-Cascaded PQCs}
Similarly, for $n$ cascaded PQCs, where each PQC takes the output of the previous one as its input, the intermediate states can be described as follows:

$$ \ket{\psi_{\text{intermediate},i}}=U_i(\theta_i)\ket{\psi_{\text{intermediate},i-1}} $$

where $i = 1, 2, \cdots, n$ and $\ket{\psi_{\text{intermediate},0}} = \ket{\psi_{\text{initial}}}$. 
The overall unitary operator of the $n$ cascaded PQCs can be obtained by composing the matrices of the individual PQCs in the correct order:

$$U_{\text{circuit}} = U_n(\theta_n)\cdots U_2(\theta_2)U_1(\theta_1)$$

The final quantum state after applying the $n$ cascaded PQCs to an initial state can be obtained by matrix-vector multiplication:

$$\left\lvert\psi_{\text{final}}\right\rangle = U_{\text{circuit}}\left\lvert\psi_{\text{initial}}\right\rangle$$

The parameters $\theta_1, \theta_2, \cdots, \theta_n$ can be optimized using classical optimization algorithms to achieve a desired quantum state or to maximize an objective function such as the expected value of a measurement outcome. The optimization problem can be written as:

    \begin{align*}
    \theta_1,\theta_2,\cdots,\theta_n &= \arg\max_{\theta_1,\theta_2,\cdots,\theta_n} \left\lvert\left\langle\psi_{\text{desired}}\right\vert U_{\text{circuit}}(\theta_1,\theta_2,\cdots,\theta_n)\left\lvert\psi_{\text{initial}}\right\rangle\right\rvert^2 \\ 
    &= \arg\max_{\theta_1,\theta_2,\cdots,\theta_n} \left\lvert\left\langle\psi_{\text{desired}}\right\vert U_n(\theta_n)\cdots U_2(\theta_2)U_1(\theta_1)\left\lvert\psi_{\text{initial}}\right\rangle\right\rvert^2 
\end{align*}
Solving this optimization problem returns the optimal set of parameters $(\theta_1,\theta_2,\cdots,\theta_n)$ that produce the desired outcome.


\subsection{Residual PQCs}
We now introduce residual blocks in the cascaded PQCs encapsulated in QNs which we call  ResQNets. 
In ResQNets, the output of the previous PQC is added to its input and fed as an input to the next PQC. The residual block is inserted to facilitate efficient information flow and improved performance. 
The primary objective of incorporating residual blocks in QNNs here is to overcome the difficulties associated with BP and thereby improve the learning process.
Furthermore, the proposed method aims to harness the strengths of both residual learning and quantum computing to tackle complex problems more effectively.

To mathematically formulate our proposed ResQNets, we start by considering the case of two PQCs, and extend the approach to the general case of cascading $n$ PQCs with $n$ residual blocks. We will refer to each PQC as $U_i$ where $i$ denotes the QN it is encapsulated in. 


\subsubsection{1-Residual Block}\label{sec:1_residual_block_theory}
\begin{figure}[h]
    \centering
    \includegraphics[scale=0.5]{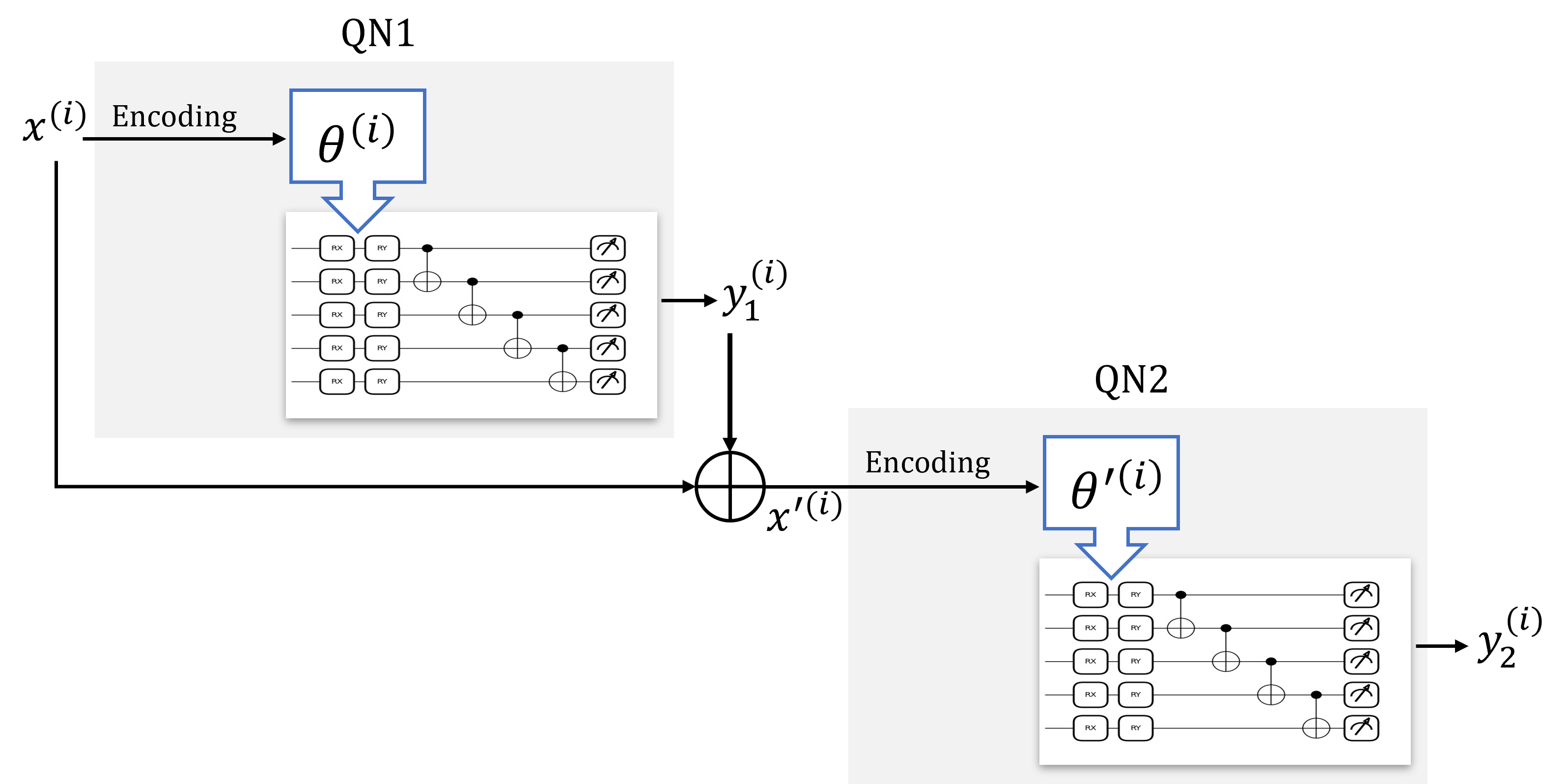}
    \caption{Illustration of residual approach in hybrid quantum neural networks}
    \label{fig:res_PQC}
\end{figure}

ResQNet with a single residual block contains a maximum of two PQCs of arbitrary depth enclosed in two separate QNs. The first QN serves as a residual block whose input is added to its output before passing it as input to the PQC in the next QN.
In the context of the NISQ era, hybrid QNNs have gained considerable traction. These models exhibit a distinctive architecture wherein the input data, characterized by its classical nature, necessitates an initial encoding process. This encoding procedure plays a vital role in preparing the data for processing on quantum computer.
It is important to note that in this paper we exclusively employ a configuration wherein classical datasets are not utilized. Instead, our approach involves the initialization of qubits in ground states and the gates in PQC are randomly parameterized prior to the training phase. Nevertheless, here we present a comprehensive mathematical framework that accommodates the broader context of hybrid systems. This formalism is especially pertinent in scenarios where classical datasets form an integral component of the computational process.
An illustrative configuration featuring a pair of Quantum Nodes (QNs), where the initial node functions as the residual block, is depicted in Figure \ref{fig:res_PQC}.
%

The two QNs will have two PQCs denoted as $U_1(\theta_1)$ and $U_2(\theta_2)$, where $\theta_1$ and $\theta_2$ are classical parameters encoded in such a way that the quantum circuit can process them.
%
The classical dataset is a set of data points, i.e., $\mathcal{D}=\{x^{(i)}\}$, where $x^{(i)}$ represents the $i^{th}$ datapoint. The next step is to encode the classical data  Each data point $x^{(i)}$, is encoded using an encoding method (e.g., angle or amplitude encoding\cite{LaRose:2020}). The angle-encoded data for the $i^{th}$ data point can be denoted as $\theta^{(i)}$. 
Each encoded data point $\theta^{(i)}$ is used as parameters for the PQC in the first quantum node (QN1). This PQC processes the encoded data and upon measurement, it generates a classical result $y_1^{(i)}$. 
The classical result $y_1^{(i)}$ from QN1 is added to the original data $x^{(i)}$ element-wise to obtain a new modified classical dataset denoted by $\mathcal{D'}=\{x^{(i)} + y_1^{(i)}\}$. 
Each data point in $\mathcal{D^{'}}$ is then encoded again to obtain a new set of encoded data denoted by $\theta^{'(i)}$ which is then used as input for a PQC in the second quantum node (QN2). 
The PQC in second QN processes this encoded data and upon measurement it generates classical result $y_2^{(i)}$ which, in case of two QNs, is the final output of the network used for cost function optimization. 

The mathematical formulation for a single residual block in two QN setting starts with encoding the classical data features into quantum space. The initial quantum state would then be:
$$ \theta^{(i)} = \ket{\psi_{\text{initial}}} =  \textit{f}_{encode}(x^{(i)}) $$
where $\textit{f}_{\text{encode}}$ is the encoding function which maps the classical input features to quantum space.
The first PQC $U_1(\theta_1)$ is applied to an initial quantum state $\left\lvert\psi_{\text{initial}}\right\rangle$ to obtain an intermediate quantum state $\left\lvert\psi_{\text{intermediate}}\right\rangle$:
$$\left\lvert\psi_{\text{intermediate}}\right\rangle = U_1(\theta_1)\left\lvert\psi_{\text{initial}}\right\rangle$$
Upon measurement the intermediate quantum state $\left\lvert\psi_{\text{intermediate}}\right\rangle$ collapses and returns the classical result:
$$ y_1 = \mathcal{M}\left\lvert\psi_{\text{intermediate}}\right\rangle $$
where $\mathcal{M}$ denotes the qubit measurement.
Now, the input of the second PQC $U_2(\theta_2)$ is not just the output of QN1 but the sum of the original input ($x^{(i)}$) and the intermediate result ($y_1$).
$$x'^{(i)} = x^{(i)} + y_1^{(i)}$$
We again have to encode the new data points obtained after addition of original input and intermediate result before passing it to the PQC in QN2. 
$$ \theta'^{(i)} = \ket{x_{\text{input}}} =  \textit{f}_{encode}(x'^{(i)}) $$
The final quantum state obtained after the second QN would be:
$$\left\lvert\psi_{\text{final}}\right\rangle = U_2(\theta_2)\ket{x_{\text{input}}} $$
After measuring the qubits in second QN, the final quantum state collapses and we get the final classical result:
$$ y_2 = \mathcal{M}\ket{\psi_{\text{final}}} $$
The parameters $\theta_1$ and $\theta_2$ can be optimized using classical optimization algorithms to achieve a desired quantum state or to maximize an objective function such as the expected value of a measurement outcome. 


\subsubsection{2-Residual blocks}
In ResQNets with two residual blocks, up to three PQCs can be incorporated within three QNs. There are three potential configurations for the residual blocks in this setup: 
\begin{enumerate}
    \item utilizing only the first QN as a residual block,
    \item combining the first two QNs to form a single residual block,
    \item utilizing both the first and second QNs individually as separate residual blocks.
\end{enumerate}

For our mathematical formulation, only the third configuration will be considered since it is the general setting for the case of two residual blocks; other configurations effectively contain a single residual block, which has already been mathematically derived in sect. \ref{sec:1_residual_block_theory}. However, we will conduct experiments examining all three configurations to determine which configuration performs the best.
Let $U_1(\theta_1)$, $U_2(\theta_2)$, and $U_3(\theta_3)$ be PQCs enclosed in three QNs, where $\theta_1$, $\theta_2$, and $\theta_3$ are the quantum-encoded classical parameters.
The initial quantum state would be obtained via encoding the classical input  features:
$$\ket{\psi_{\text{initial}}} = \textit{f}_{encode}(x^{(i)})$$
The first PQC $U_1(\theta_1)$ takes the initial quantum state $\left\lvert\psi_{\text{initial}}\right\rangle$ as its input and produces an intermediate quantum state $\left\lvert\psi_{\text{intermediate}}\right\rangle$:
$$\ket{\psi_{\text{intermediate}}} = U_1(\theta_1)\ket{\psi_{\text{initial}}}$$
The $\ket{\psi_{\text{intermediate}}}$ is the output of $U_1(\theta_1)$ before the measurement. Upon measuring the qubits the quantum state $\ket{\psi_{\text{intermediate}}}$ collapses and produces a classical result:
$$y_1 = \mathcal{M}\left\lvert\psi_{\text{intermediate}}\right\rangle$$
Before passing the output of QN1 ($y_1$) as input to QN2, it is first added element-wise with the original input $x^{(i)}$. Since, both $y_1$ and $x^{(i)}$ are classical values, therefore an encoding function is again applied in order for the PQC in QN2 to process them:
$$x'^{(i)} = x^{(i)} + y_1^{(i)}$$
$$\ket{\psi_{\text{input}}^{(1)}} = \textit{f}_{encode}(x'^{(i)})$$
The encoded data is then passed to second PQC, yielding another intermediate quantum state $\left\lvert\psi_{\text{intermediate}}'\right\rangle$:
$$\ket{\psi_{\text{intermediate}}'} = U_2(\theta_2)\ket{\psi_{\text{input}}^{(1)}}$$
The quantum state $\ket{\psi_{\text{intermediate}}'}$ is the result before measurement and upon measuring the PQC in QN2, we get the classical result:
$$ y_2 = \mathcal{M}\ket{\psi_{\text{intermediate}}'}$$
Finally, the third PQC $U_3(\theta_3)$ takes the sum of output of QN1 (which also is the input of QN2) and output of QN2 as input and produces the final quantum state $\left\lvert\psi_{\text{final}}\right\rangle$: 
$$x''^{(i)} = y_1^{(i)} + y_2^{(i)}$$
Since $x''^{(i)}$ is again classical, so it has to encoded before passing it to the $U_3(\theta_3)$:
$$ \ket{\psi_{\text{input}}^{(2)}} = \textit{f}_{encode}(x''^{(i)})$$
The final quantum state obtained after the action of $U_3(\theta_3)$ on $x_2$ will be:
$$ \ket{\psi_{\text{final}}} = U_3(\theta_3)\ket{\psi_{\text{input}}^{(2)}}$$
The final quantum state collapses into a classical vector after measurement, which will be the final result of the network used for further optimization. 
$$ y_3 = \mathcal{M}\ket{\psi_{\text{final}}}$$
The overall unitary operator of the three cascaded PQCs can be obtained by composing the matrices of the individual PQCs in the correct order:
$$U_{\text{circuit}} = U_3(\theta_3)U_2(\theta_2)U_1(\theta_1)$$
The final quantum state after applying the three cascaded PQCs to an initial state can be obtained by matrix-vector multiplication:
$$\left\lvert\psi_{\text{final}}\right\rangle = U_{\text{circuit}}\left\lvert\psi_{\text{initial}}\right\rangle = U_3(\theta_3)U_2(\theta_2)U_1(\theta_1)\left\lvert\psi_{\text{initial}}\right\rangle$$
The same procedure can be extended for $n$-residual blocks with different residual configurations. 

Given a set of $n$ PQCs, $U_1(\theta_1), U_2(\theta_2),\dots, U_n(\theta_n)$ and an initial quantum state $\left\lvert\psi_{\text{initial}}\right\rangle$, the objective is to find the set of parameters $\boldsymbol{\theta} = {\theta_1, \theta_2,\dots, \theta_n}$ that maximizes (or minimizes) some cost function $C(\boldsymbol{\theta})$ associated with the final quantum state $\left\lvert\psi_{\text{final}}\right\rangle$ produced by the cascaded PQCs. The optimization problem can be formulated as:

$$\boldsymbol{\theta}^\ast = \arg\max_{\boldsymbol{\theta}} C(\boldsymbol{\theta})$$

\noindent or

$$\boldsymbol{\theta}^\ast = \arg\min_{\boldsymbol{\theta}} C(\boldsymbol{\theta})$$

\noindent where $\boldsymbol{\theta}^\ast$ represents the optimal set of parameters that maximizes (or minimizes) the cost function. The cost function $C(\boldsymbol{\theta})$ can be defined based on the desired behavior of the quantum circuit and can be calculated from the measurement result of final quantum state $\ket{\psi_{\text{final}}}$.


\section{Methodology}\label{sec:methodology}
In classical NNs, residual neural networks (ResNets) were proposed to overcome the problem of vanishing gradients and were very useful for enabling deep learning in classical machine learning. 
In this paper, we propose a Residual Quantum Neural Networks (ResQNets), to enable deep learning in QNNs by mitigating the effect of BP as a function of the number of layers. 

\begin{figure*}[h!]
     \centering
    \begin{subfigure}[b]{0.75\textwidth}
         \centering
    \includegraphics[scale=0.6]{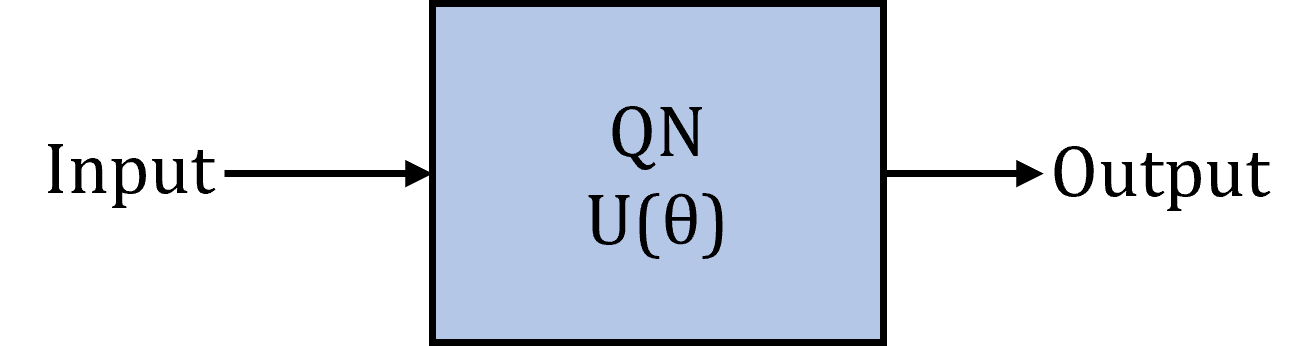}
         \caption{}
         \label{fig:SPN_arch}
     \end{subfigure}
        \hfil
     \begin{subfigure}[b]{0.75\textwidth}
         \centering
    \includegraphics[scale=0.6]{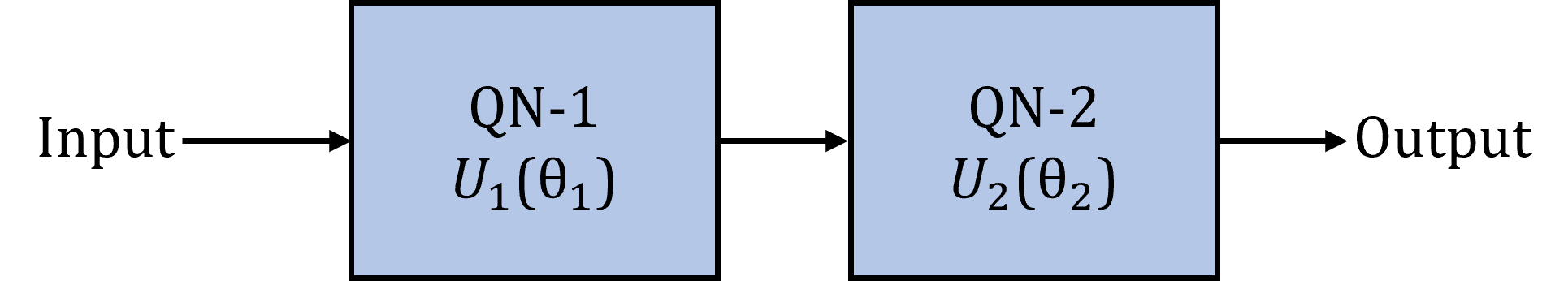}
         \caption{}
         \label{fig:PN_arch}
     \end{subfigure}
        \hfil
        \begin{subfigure}[b]{0.75\textwidth}
        \hspace{-1cm}
        \includegraphics[scale=0.6]{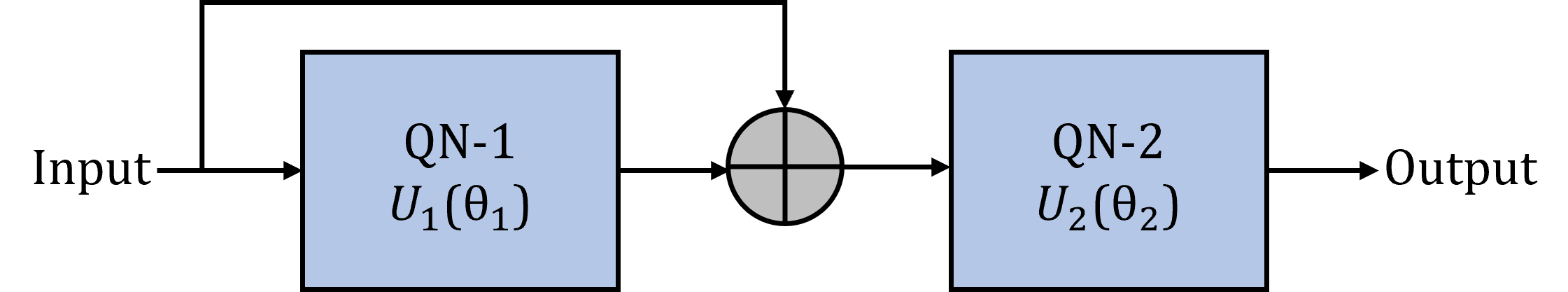}
         \caption{}
         \label{fig:RN_arch}
     \end{subfigure}
     \captionsetup{hypcap=false}
    \caption{QNN architecture used in this paper (a) Simple PlainQNet (b) PlainQNet and (c) ResQNet. \textcolor{blue}{The internal architure and working of QN is shown Figure \ref{fig:res_PQC}}.}  
    \label{fig:PN_RN_archs}
    
\end{figure*}

The conventional approach to constructing QNNs contains an arbitrarily deep PQC, which takes some input and yields some output. Such an architecture typically has a single QN, as depicted in Fig. \ref{fig:SPN_arch}. In this paper, we refer to this traditional QNN architecture as ``Simple PlainQNet''.

To construct our proposed ResQNets, we need to further split the traditional QNN architecture into two QNs, where every QN contains arbitrary deep quantum layers.
Since our proposed ResQNets contain at least two QNs and the traditional way of constructing QNNs contains a single QN, we construct a slightly modified version of simple PlainQNet, which we call ``PlainQNet'' and includes two or more QNs, with each QN containing PQCs of arbitrary depth, as shown in Fig. \ref{fig:PN_arch}. In PlainQNets, the output of the previous QN is fed to the next QN. 
The purpose of constructing PlainQNet is to have a fair comparison with our proposed ResQNets because ResQNets need two or more QNs to work. An example of ResQNet architecture with two QNs is shown in Fig. \ref{fig:RN_arch}. 
The PlainQNet architecture is similar to general QNN split into two QNs, whereas in the case of ResQNet, the first QN serves as the residual block, i.e., the input of the first QN is added to its output and then passed as input to the second QN.

It should be noted that ResQNets can comprise multiple QNs with various arrangements of residual blocks. For instance, the ResNet from Fig. \ref{fig:RN_arch} can be extended to have three QNs, in which case three potential configurations can be employed. These include having the first and second QNs acting as individual residual blocks, combining the first and second QNs to serve as a single residual block, and only the first QN functioning as the residual block. The possibility of these three configurations has been taken into consideration. We also consider the case of three QNs with these configurations. 


\subsection{Quantum Layers Design}
 
\begin{figure*}[h!]
    \centering
    \includegraphics[scale=0.4]{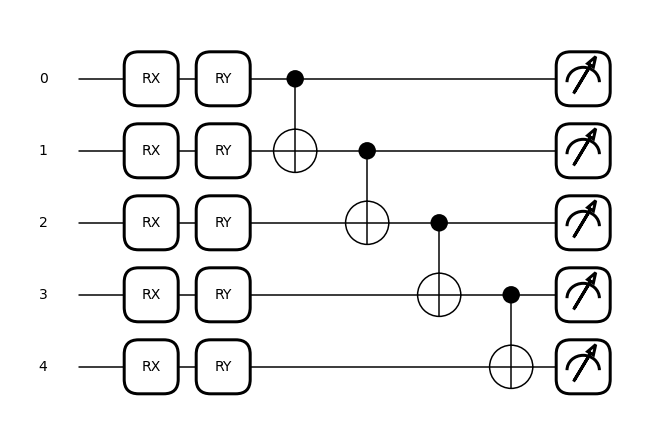}
    \captionsetup{hypcap=false}
    \caption{Quantum Layers Design}
        \label{fig:Qlayers}
\end{figure*}

For the design of quantum layers, we use a periodic structure containing two single-qubit unitaries ($RX$ and $RY$) per qubit. These unitaries are randomly initialized in the range [$0,\pi$]. Furthermore, a two-qubit gate, i.e., $CNOT$-gate is used to entangle qubits, and every qubit is entangled with its neighboring qubit. Fig. \ref{fig:Qlayers} shows the example design of the quantum layers we used (5 qubits). All the QNs in our experiments have the same quantum layers desgin.       

\subsection{Depth of Quantum Layers}
The impact of the quantum layer depth in examining the existence of BP in the cost function landscape of a QNN is significant.
Effective depth (the longest path within the quantum circuit until the measurement) is crucial in this regard. 
For convenience, We introduce two depth parameters: layer depth ($D_L$) and effective depth ($D_E$). The layer depth $D_L$ refers to the combined number of repetitions of the quantum layer illustrated in Fig. \ref{fig:Qlayers} in both QNs, while the effective depth $D_E$ represents the overall depth.   
For our quantum layers design, the following equation can be used to calculate the effective depth. 

\begin{equation}\label{eq:eff_depth}
     \textit{Total Effective Depth} = D_E = 4 \times D_L + k
\end{equation} 

\noindent where $k = 2, 3,4,5\ldots$ for  $5,6,7,8\ldots$ qubits, respectively. Since the quantum layers are split into two separate QNs, and the depth per QN can be crucial to achieving better performance, it is important to calculate $D_E$ of each QN individually and then add them to obtain the final $D_E$.
Failure to calculate the depth in each QN separately could result in an effective depth different from the sum of the effective depths of each QN, i.e., $D_L/QN1 + D_L/QN2 \neq D_E$.
For example, with $D_L=2$, the total effective depth would be 10 without considering the splitting into two QNs. However, if $D_L$ is split into two QNs with $D_L/QN=1$, the effective depth would be 12.
A modified version of Eq. (\ref{eq:eff_depth}) should be used to calculate the $D_E$ per QN, as described below.

\begin{equation}\label{eq:eff_depth/QN}
\textit{Effective Depth per QN} = D_E/QN = 4 \times D_L/QN + k 
\end{equation}
\subsection{Depth Distribution per QN}
As previously discussed, ResQNets and PlainQNets consist of multiple QNs, which results in different depth splits for a given depth of quantum layers. According to the definition of BP, the gradient vanishes as a function of the number of qubits; hence, we fix the depth of quantum layers to $D_L=6$, and only vary the number of qubits.
Table \ref{tab:depthperQN} summarizes the different depth per QN combinations for $D_L=6$, and all these depth combinations are tested for different numbers of qubits. Column 3 of Table \ref{tab:depthperQN} represents the depth split in the form of ordered pairs (we refer to this form in the rest of the paper whenever we discuss depth split per QN). For instance, $(1,5)$ denotes $D_L=1$ in the first QN and $D_L=5$ in the second QN.
The depth per QN combination can be extended to more than two QNs in a similar manner.

\begin{table}[htp]
    \centering
    \caption{Depth combinations per QN}
    \begin{tabular}{|p{3.5cm}|p{3.5cm}|p{3.5cm}|}
    \hline
         \hfil$D_L$ in QN-1 & \hfil$D_L$ in QN-2 &\hfil in-text representation\\
         \hline
         \hfil 1 &\hfil 5 &\hfil (1,5)\\
         \hline
         \hfil 5 & \hfil 1 &\hfil (5,1)\\
         \hline
         \hfil 2 &\hfil 4 &\hfil (2,4)\\
         \hline
          \hfil 4 &\hfil 2 &\hfil (4,2)\\
          \hline
           \hfil 3 &\hfil 3 &\hfil (3,3)\\

         \hline
    \end{tabular}
    
    \label{tab:depthperQN}
\end{table}

\subsection{Cost Function Definition}
For training our proposed ResQNet, we consider a simple example of learning the identity gate. In such a scenario a natural cost function would be the difference of $1$ minus the probability of measuring an all-zero state, which can be described by the following equation.

$$  C = \bra{\psi(\theta)} (I-\ket{0}\bra{0}) \ket{\psi(\theta)} = 1- p_{\ket{0}} 
$$

We consider the global cost function setting, i.e., we measure all the qubits in the network. Therefore, the above cost function definition will be applied across all the qubits according to the following equation.

\begin{equation}\label{eq:CF_eq}
    C = \bra{\psi(\theta)} (I-\ket{00\ldots0}\bra{00\ldots0}) \ket{\psi(\theta)} = 1- p_{\ket{00\ldots0}} 
\end{equation}

For cost function optimization, we use Adam optimizer (with a stepsize of 0.1), which is a gradient-based optimization method for optimization problems. The Adam optimizer updates the parameters of a model iteratively based on the gradient of the loss function with respect to the parameters.
The Adam optimizer uses an exponentially decaying average of the first and second moments of the gradients to adapt the learning rate for each parameter. Let $g_{t}$ be the gradient of the loss function with respect to the parameters at iteration $t$. The first moment, $m_{t}$, and the second moment, $v_{t}$, are computed as follows:

\begin{align*}
m_{t} &= \beta_{1} m_{t-1} + (1 - \beta_{1}) g_{t} \\
v_{t} &= \beta_{2} v_{t-1} + (1 - \beta_{2}) g_{t}^{2}
\end{align*}

where $\beta_{1}$ and $\beta_{2}$ are the decay rates for the first and second moments, respectively.
The bias-corrected first moment and second moment are then computed as:

\begin{align*}
\hat{m}_{t} &= \frac{m_{t}}{1 - \beta_{1}^{t}} \\
\hat{v}_{t} &= \frac{v_{t}}{1 - \beta_{2}^{t}}
\end{align*}

Finally, the parameters are updated using the following equation:

\begin{align*}
\theta_{t+1} = \theta_{t} - \frac{\alpha}{\sqrt{\hat{v}_{t}} + \epsilon} \hat{m}_{t}
\end{align*}

where $\alpha$ is the learning rate and $\epsilon$ is a small constant to prevent division by zero.


\section{Results and Discussion}\label{sec:results}
In order to investigate the issue of BP in both PlainQNets and ResQNets, we maintain a constant depth of quantum layers, $D_L=6$, which comprises $100$ quantum gates and $60$ parameters. 
The quantum layer depth distribution is varied among different combinations, as discussed in Table \ref{tab:depthperQN}. The $D_E$ per QN can then be calculated using Eq. (\ref{eq:eff_depth/QN}).
The performance of both networks is evaluated by comparing their cost function landscapes and training results for the problem specified in Eq. (\ref{eq:CF_eq}).

\subsection{PlainQNet and Simple PlainQNet}

In this paper, the construction of the proposed ResQNets involves the incorporation of a minimum of two QNs, whereas traditionally QNNs development entails the use of a single QN (referred to as ``simple PlainQNets in this paper).
To ensure a fair performance comparison of QNNs with no residual connections and our proposed ResQNets, we modify the architecture of simple PlainQNets by dividing it into two QNs (referred to as ``PlainQNets" in this paper). 
This architectural modification is primarily aimed to have a similar architecture of PlainQNets (QNNs with no residual connection) and ResQNets before comparing their performance. 


Given the modification introduced to the conventional QNN architecture, as stated above, it is necessary to comparatively analyze the performance of the unaltered simple PlainQNets and the adapted PlainQNets. This preliminary comparison aims to identify any potential consequences arising from the structural modification. If this architectural modification results in minimal disruptions to performance, it would establish a basis for conducting a subsequent comparative analysis between PlainQNets and ResQNets with confidence.

\begin{figure*}[h!]
     
        \begin{subfigure}[b]{0.2\textwidth}
         \hspace*{-2.3cm}
        \includegraphics[scale=0.27]{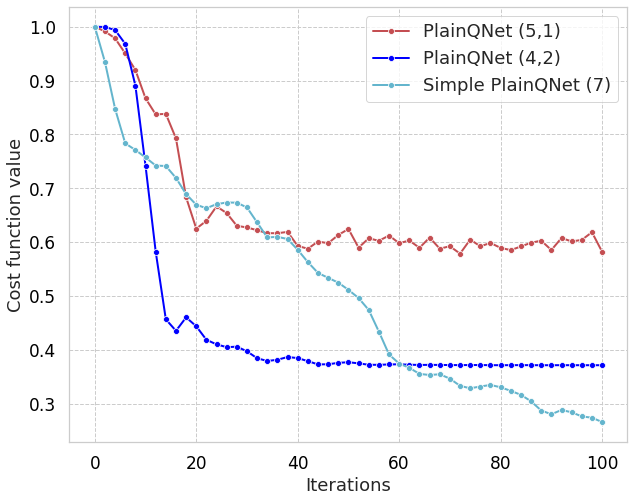}
         \caption{}
         \label{fig:training_PNvsSPN_6Q}
     \end{subfigure}
       \hfil\quad
\begin{subfigure}[b]{0.2\textwidth}
    \hspace*{-1.2cm}
         \includegraphics[scale=0.27]{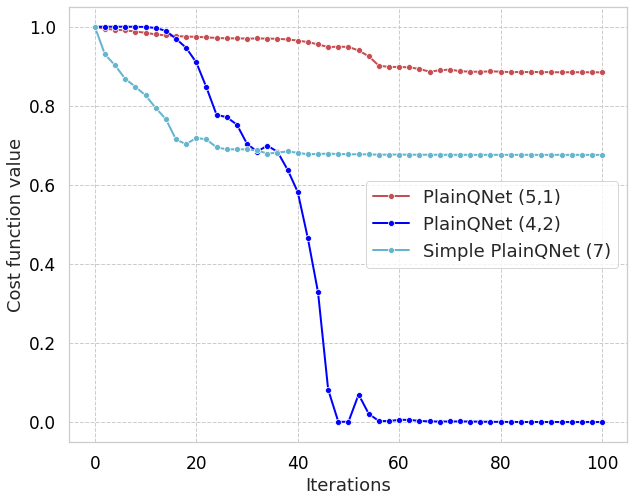}
         \caption{}
        \label{fig:training_PNvsSPN_7Q}
     \end{subfigure}
     \captionsetup{hypcap=false}
    \caption{Cost vs. iterations of PlainQNets and simple PlainQNets (a) for $6$ qubits (b) for $7$ qubits. The parentheses denote the $D_L$ per QN.}  
    \label{fig:training_PN_SPN_6Q_7Q}
    
\end{figure*}

The simple PlainQNets and PlainQNets are compared for 6-qubit and 7-qubit quantum layers with a constant depth of $D_L=6$. In the case of PlainQNets, the depth distribution per QN can vary, but we use the depth combinations of $(5,1)$ and $(4,2)$, where the first entry represents the depth of the first QN and the second entry represents the depth of the second QN, as shown in Table \ref{tab:depthperQN}. 
We choose deeper quantum layers on the first QN and relatively shallow depth on the second QN primarily because such a configuration of depths per QN leads to a better performance, which will be discussed in more detail in the subsequent sections.
For 6-qubit quantum layers, the effective depth ($D_E$) for PlainQNets for both depth combinations mentioned above is $30$ (as defined in Eq. (\ref{eq:eff_depth/QN})). The closest possible $D_E$ for simple PlainQNets using the quantum layers considered in this paper (shown in Fig. \ref{fig:Qlayers}) is $31$ with an overall $D_L$ of $7$ (as defined in Eq. (\ref{eq:eff_depth})), which was used in the comparison.
Similarly, for 7-qubit quantum layers, the $D_E$ for PlainQNets is $32$ for both depth combinations per QN. The closest $D_E$ in the case of simple PlainQNets is obtained for $D_L=7$.

Both PlainQNets and simple PlainQNets are then trained for the problem specified in Eq. (\ref{eq:CF_eq}). The training results are displayed in Fig. \ref{fig:training_PN_SPN_6Q_7Q}. It can be observed that for 6-qubit layers, both PlainQNets and simple PlainQNets exhibit comparable performance. However, when the number of qubits increases to 7, the performance of simple PlainQNets decreases significantly due to BP, while PlainQNets improves. Based on these observations, we can infer that it is appropriate to compare the performance of PlainQNets with that of our proposed ResQNets. Hence, for the remainder of the paper, we will compare the performance of PlainQNets, which are QNNs containing two (or more) QNs, with that of ResQNets.
\subsection{ResQNet with shallow width quantum layers} \label{sec:RN_shallow}
In this section, we perform a comparative analysis of the incidence of BP in both PlainQNets and ResQNets. Both PlainQNets and ResQNets consist of two QNs, with a maximum of one residual block in the case of ResQNets. 
To facilitate a fair comparison, we consider shallow depth quantum layers with $D_L=6$ and incrementally vary the number of qubits from 6 to 10.


\subsubsection{6-Qubit Circuit}
\begin{figure*}[h!]
\hspace{-1cm}
\begin{multicols}{5}
\hspace*{-0.65cm}
    \includegraphics[scale=0.21]{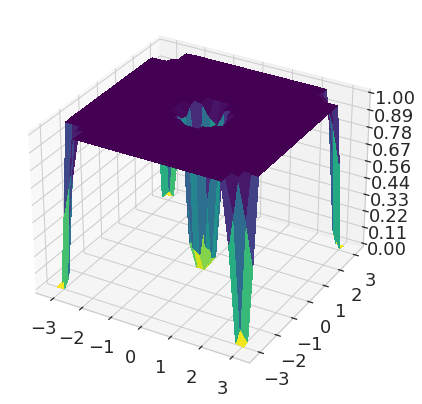}\par
    \caption*{(1,5)}\label{fig:6Q_PN_15}
    \hspace*{-0.65cm}
    \includegraphics[scale=0.21]{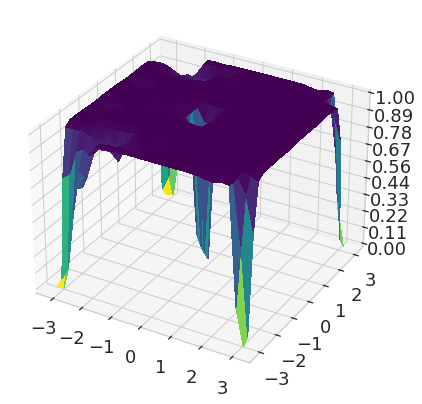}\par
    \caption*{(5,1)}\label{fig:6Q_PN_51}
    \hspace*{-0.65cm}
     \includegraphics[scale=0.21]{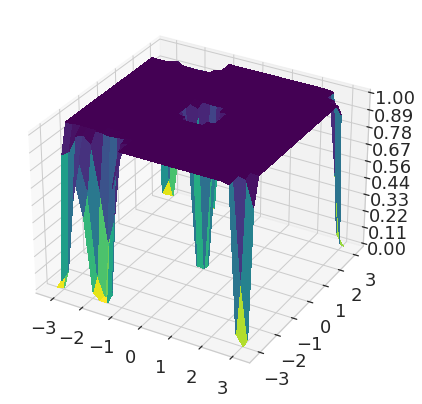}\par
     \caption*{(2,4)}\label{fig:6Q_PN_24}
     \hspace*{-0.65cm}
     \includegraphics[scale=0.21]{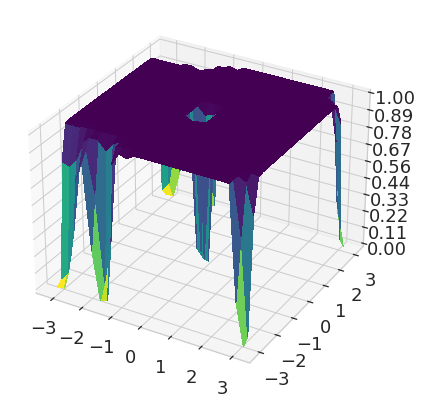}\par
     \caption*{(4,2)}\label{fig:6Q_PN_42}
     \hspace*{-0.65cm}
     \includegraphics[scale=0.21]{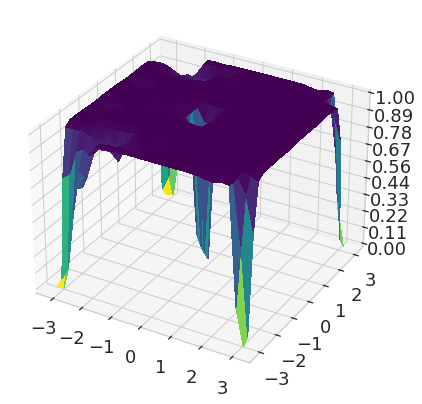}\par
     \caption*{(3,3)}\label{fig:6Q_PN_33}
\end{multicols}

\begin{multicols}{5}
\hspace*{-0.65cm}
    \includegraphics[scale=0.21]{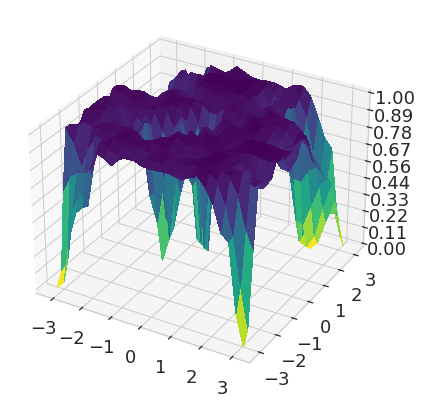}\par
    \caption*{(1,5)}
    \label{fig:6Q_RN_1_5}
\hspace*{-0.65cm}
    \includegraphics[scale=0.21]{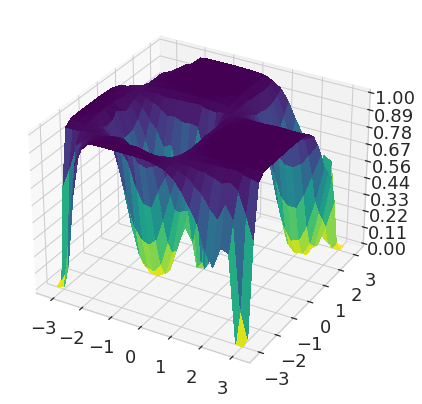}\par
    \caption*{(5,1)}
    \label{fig:6Q_RN_5_1}
\hspace*{-0.65cm}
     \includegraphics[scale=0.21]{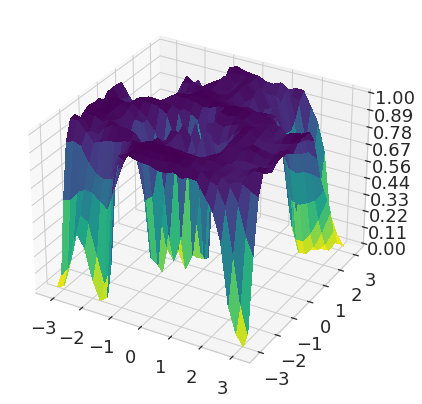}\par
     \caption*{(2,4)}
     \label{fig:6Q_RN_2_4}
\hspace*{-0.65cm}
     \includegraphics[scale=0.21]{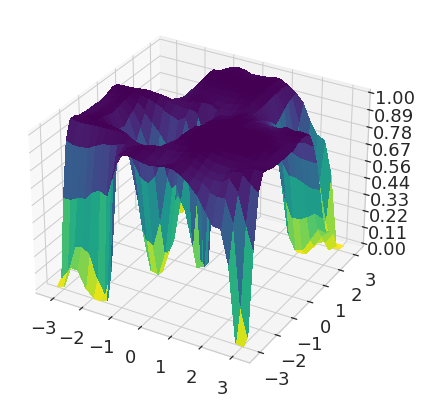}\par
     \caption*{(4,2)}
     \label{fig:6Q_RN_4_2}
\hspace*{-0.65cm}
      \includegraphics[scale=0.21]{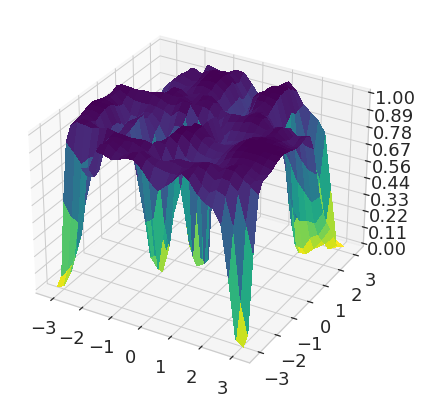}\par
      \caption*{(3,3)}
      \label{fig:6Q_RN_3_3}
\end{multicols}
\captionsetup{hypcap=false}
\caption{Cost function landscapes of PlainQNet (upper panel) and ResQNet (lower panel) for $6$ Qubits. The parentheses denotes the $D_L$ per QN.}
\label{fig:cf_landscape_PN_RN_Q=6}
\end{figure*}

In this setting, we experiment with a total of $6$ qubits.
The cost function landscapes for both PlainQNet and ResQNet were analyzed and compared, as shown in Fig. \ref{fig:cf_landscape_PN_RN_Q=6}. 
The results demonstrate that a significant portion of the cost function landscapes of the PlainQNet for almost all the depth combinations are flat and have a narrow region containing the global minimum. 
On the other hand, the cost function landscapes of ResQNets are less flat and have a wider region containing the global minimum, which makes ResQNet more suitable for optimization. 

\begin{figure}[h!]
     
        \begin{subfigure}[b]{0.2\textwidth}
         \hspace*{-2.3cm}
        \includegraphics[scale=0.27]{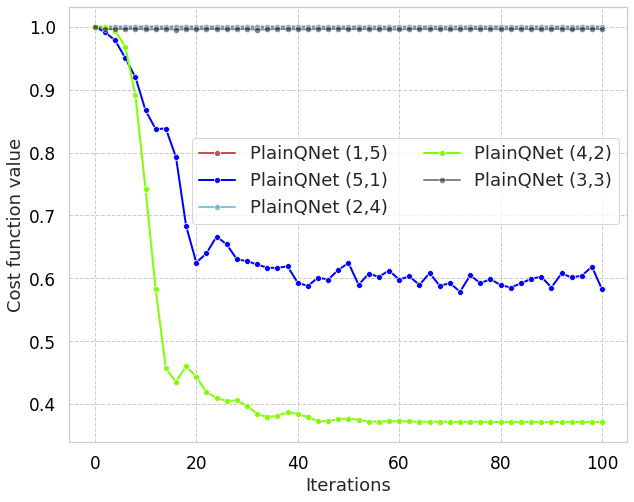}
         \caption{}
         \label{fig:training_PN_6Q}
     \end{subfigure}
       \hfil\quad
   \begin{subfigure}[b]{0.2\textwidth}
         \hspace*{-1.2cm}
         \includegraphics[scale=0.27]{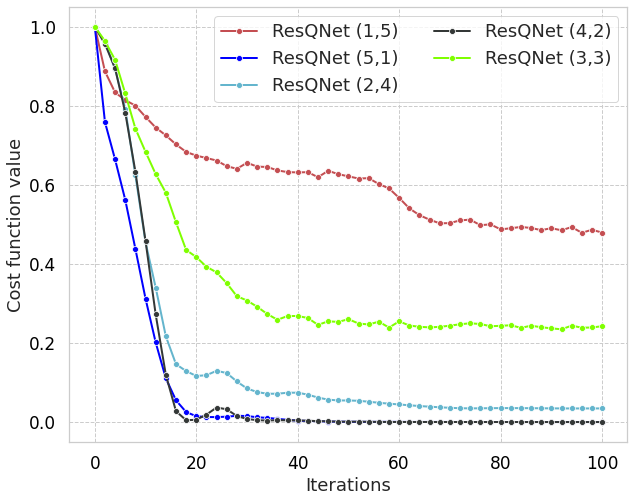}
         \caption{}
       
     \end{subfigure}
     \captionsetup{hypcap=false}
    \caption{Cost vs. iterations of (a) PlainQNets (b) and ResQNets for $6$ qubits. The parentheses denote the $D_L$ per QN. }  
    \label{fig:training_PN_RN_6Q}
    
\end{figure}

The training of PlainQNets and ResQNets was performed for the problem defined in Eq. (\ref{eq:CF_eq}). The results of the training are depicted in Fig. \ref{fig:training_PN_RN_6Q}.
When the depth of the second QN is equal to or greater than the depth of the first QN, it was observed that the PlainQNets do not undergo successful training. This can be attributed to the flat cost function landscape, i.e., the BP, as depicted in Fig. \ref{fig:cf_landscape_PN_RN_Q=6}.
For the similar depth distribution per QN (depth in second QN $\geq$ depth in first QN), the ResQNets were observed to effectively undergo training. However, they struggled to reach an optimal solution due to the presence of multiple local minima in their cost function landscape.
In instances where the depth of the first QN is greater than the second QN, both PlainQNets and ResQNets underwent successful training, but ResQNets outperformed PlainQNets. 

\subsubsection{8-Qubit Ciruit}

\begin{figure}[h!]
\begin{multicols}{5}
\hspace*{-0.5cm}
    \includegraphics[scale=0.21]{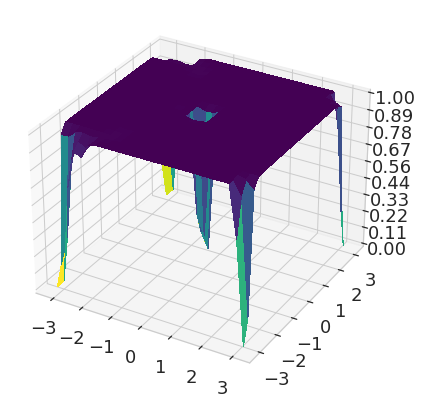}\par
    \caption*{(1,5)}\label{fig:8Q_PN_15}
    \hspace*{-0.5cm}
    \includegraphics[scale=0.21]{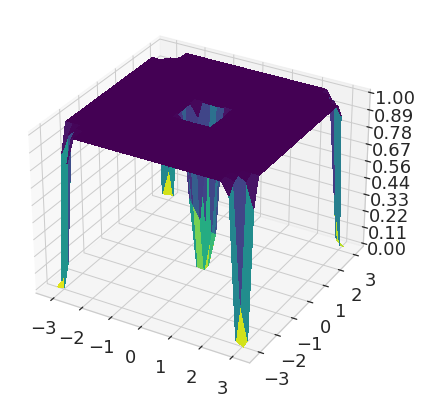}\par
    \caption*{(5,1)}\label{fig:8Q_PN_51}
    \hspace*{-0.5cm}
     \includegraphics[scale=0.21]{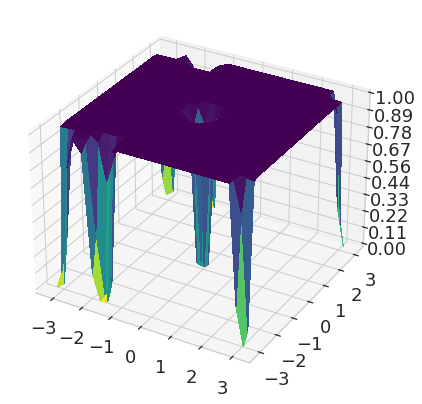}\par
     \caption*{(2,4)}\label{fig:8Q_PN_24}
     \hspace*{-0.5cm}
     \includegraphics[scale=0.21]{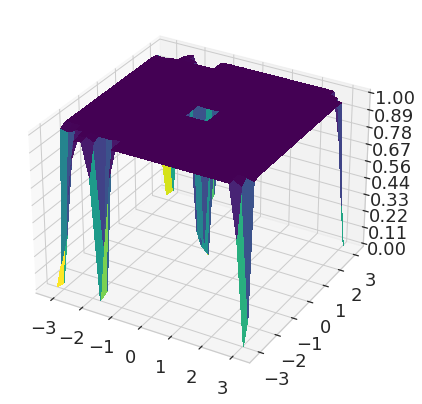}\par
     \caption*{(4,2)}\label{fig:8Q_PN_42}
     \hspace*{-0.5cm}
     \includegraphics[scale=0.21]{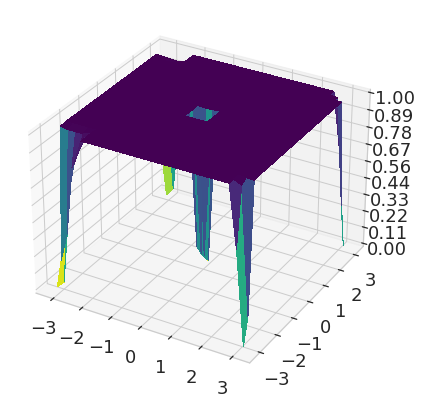}\par
     \caption*{(3,3)}\label{fig:8Q_PN_33}
\end{multicols}

\begin{multicols}{5}
    \hspace*{-0.5cm}
    \includegraphics[scale=0.21]{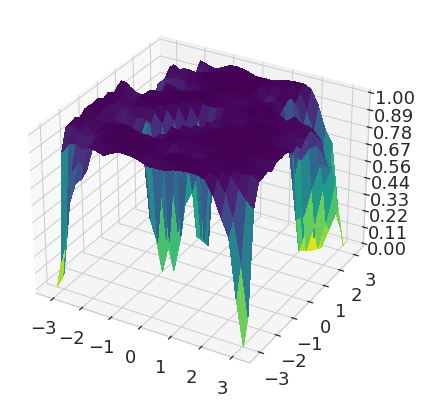}\par
    \caption*{(1,5)}\label{fig:8Q_RN_1_5}
    \hspace*{-0.5cm}
    \includegraphics[scale=0.21]{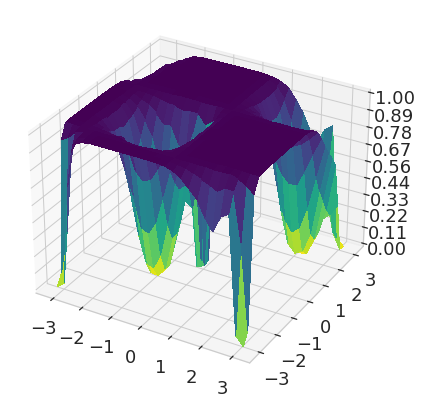}\par
    \caption*{(5,1)}\label{fig:8Q_RN_5_1}
    \hspace*{-0.5cm}
     \includegraphics[scale=0.21]{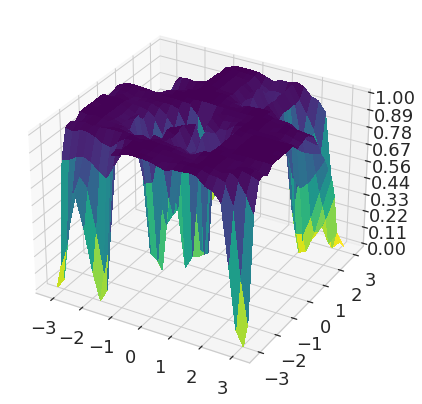}\par
     \caption*{(2,4)}\label{fig:8Q_RN_2_4}
     \hspace*{-0.5cm}
     \includegraphics[scale=0.21]{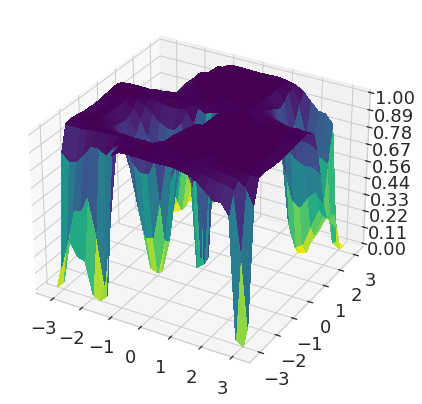}\par
     \caption*{(4,2)}\label{fig:8Q_RN_4_2}
\hspace*{-0.5cm}
      \includegraphics[scale=0.21]{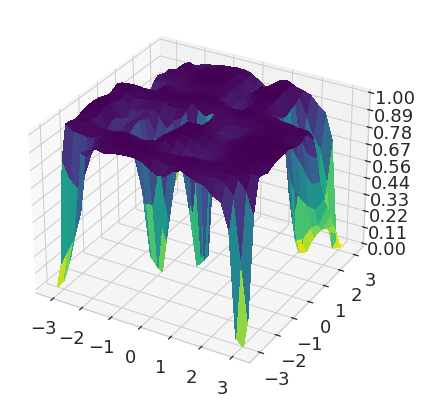}\par
      \caption*{(3,3)}\label{fig:8Q_RN_3_3}
\end{multicols}
\captionsetup{hypcap=false}
\caption{Cost function landscapes of PlainQNet (upper panel) and ResQNet (lower panel) for $8$ Qubits. The parentheses denote the $D_L$ per QN.}
\label{fig:cf_landscape_PN_RN_Q=8}
\end{figure}

We now conduct experiments on both PlainQNets and ResQNets with 8-qubit layers, and examine the cost function landscapes of both PlainQNets and our proposed ResQNets. The overall layer depth is set to 6, and all depth combinations are analyzed. The results presented in Fig. \ref{fig:cf_landscape_PN_RN_Q=8}, reveal that approximately 90\% of the cost function landscape for PlainQNets remains flat irrespective of the depth distribution per QN, making them unsuitable for optimization. In contrast, the cost function landscapes of ResQNets are still not flat for all the depth combinations, and thus are more favorable for optimization.

\begin{figure}[h!]
     \begin{subfigure}[b]{0.2\textwidth}
         \hspace*{-2.3cm}
        \includegraphics[scale=0.27]{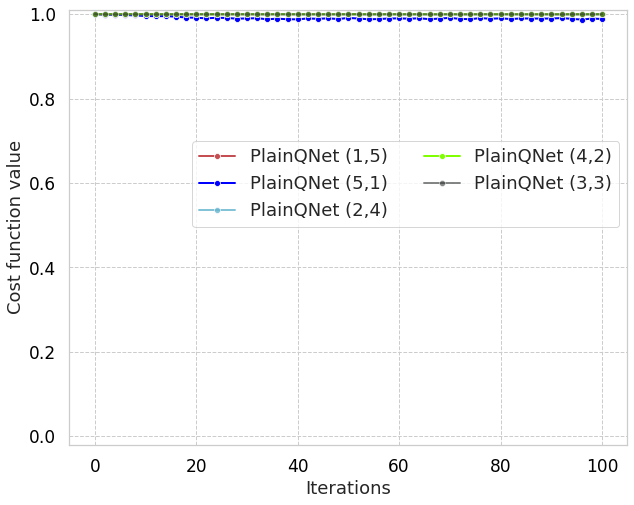}
         \caption{}
         \label{fig:training_PN_8Q}
     \end{subfigure}
       \hfil\quad
    \begin{subfigure}[b]{0.2\textwidth}
         \hspace*{-1.2cm}
         \includegraphics[scale=0.27]{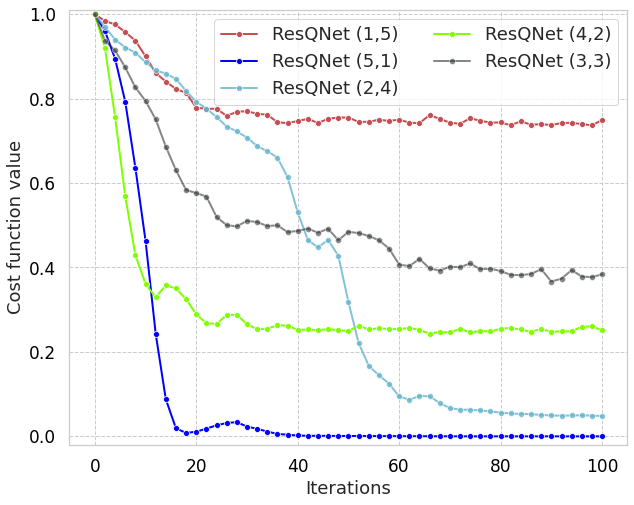}
         \caption{}
         \label{fig:training_RN_8Q}
     \end{subfigure}
     \captionsetup{hypcap=false}
    \caption{Cost vs. iterations of (a) PlainQNets and (b) ResQNets for $8$ qubits. The parentheses denote the $D_L$ per QN.}  
    \label{fig:training_PN_RN_8Q}
    
\end{figure}

We conduct training experiments for both PlainQNets and ResQNets with 8 qubit quantum layers to solve the problem defined in Eq. (\ref{eq:CF_eq}). 
The training results are presented in Fig. \ref{fig:training_PN_RN_8Q}, which show that as we increase the number of qubits from 6 to 8, the PlainQNets get trapped in the flat cost function landscape (i.e., BP), for all the depth combinations per QN and fail to train effectively for the specified problem.

On the other hand, the ResQNets demonstrate successful training across all the depth combinations, surpassing the performance of PlainQNets.
Notice that ResQNets exhibit superior learning outcomes when the depth of the first QN is much greater than that of the second QN ($D_E in QN1>>>>D_E in QN2$), such as in the case of $(5,1)$. This is because in such scenarios the cost function landscape has fewer and wider regions leading to the global minimum. Conversely, when the depth of the second QN is equal to or greater than that of the first QN, the cost function landscape is characterized by multiple local minima, making it less suitable for optimization as the optimizer becomes trapped in local minima.
This phenomenon can be attributed to the presence of residual blocks in ResQNets. In the case of two QNs, a residual connection is introduced only after the first block. This helps in mitigating the issue of BP. However, if the second QN is deep enough, it can still result in BP. In such scenarios, the cost function landscape still contains multiple local minima and fewer paths to reach the global minimum, which makes the optimization process more prone to becoming stuck in a local minimum. Despite this, ResQNets still demonstrate superior training performance compared to PlainQNets.

\subsubsection{10-Qubit Circuit}

\begin{figure*}[h]
\begin{multicols}{5}
\hspace*{-0.5cm}
    \includegraphics[scale=0.21]{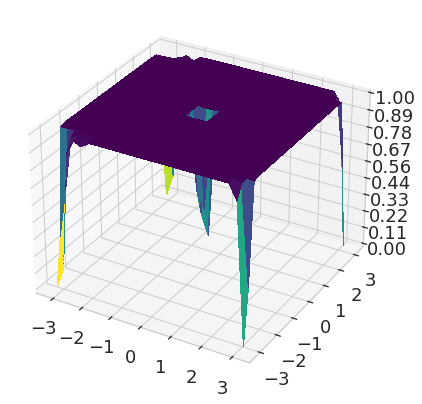}\par
    \caption*{(1,5)}\label{fig:10Q_PN_15}
    \hspace*{-0.5cm}
    \includegraphics[scale=0.21]{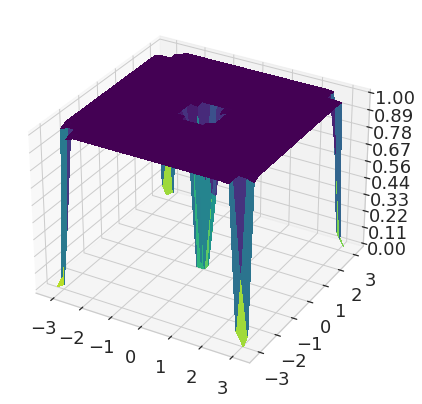}\par
    \caption*{(5,1)}\label{fig:10Q_PN_51}
    \hspace*{-0.5cm}
     \includegraphics[scale=0.21]{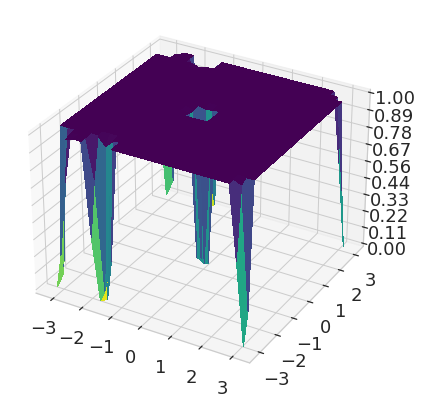}\par
     \caption*{(2,4)}\label{fig:10Q_PN_24}
     \hspace*{-0.5cm}
     \includegraphics[scale=0.21]{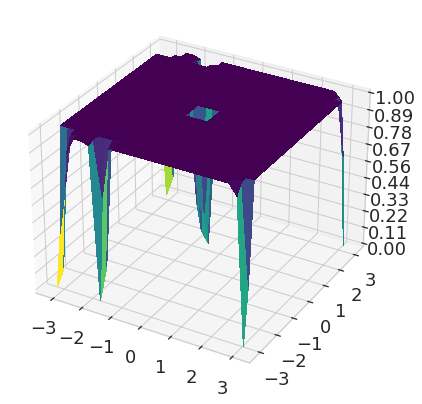}\par
     \caption*{(4,2)}\label{fig:10Q_PN_42}
     \hspace*{-0.5cm}
     \includegraphics[scale=0.21]{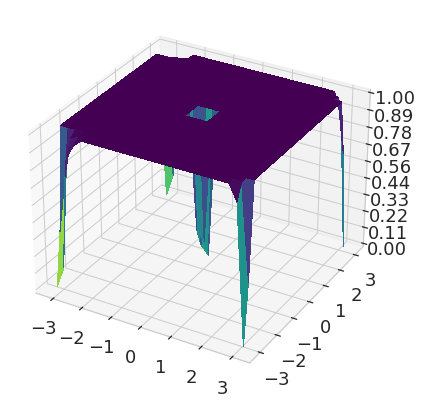}\par
     \caption*{(3,3)}\label{fig:10Q_PN_33}
\end{multicols}

\begin{multicols}{5}
    \hspace*{-0.5cm}
    \includegraphics[scale=0.21]{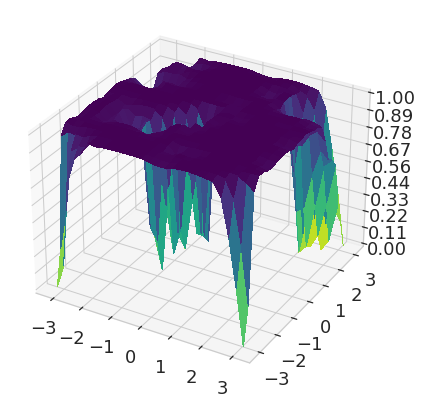}\par
    \caption*{(1,5)}\label{fig:10Q_RN_1_5}
    \hspace*{-0.5cm}
    \includegraphics[scale=0.21]{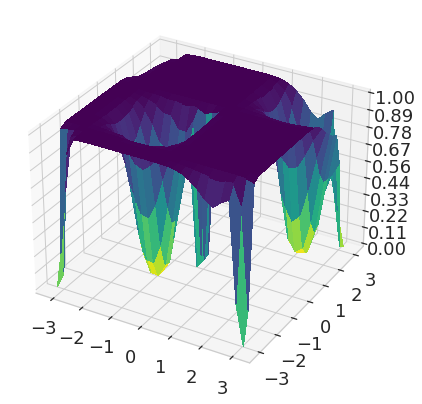}\par
    \caption*{(5,1)}\label{fig:10Q_RN_5_1}
    \hspace*{-0.5cm}
     \includegraphics[scale=0.21]{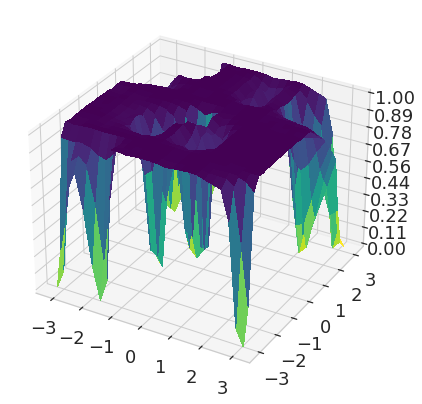}\par
     \caption*{(2,4)}\label{fig:10Q_RN_2_4}
     \hspace*{-0.5cm}
     \includegraphics[scale=0.21]{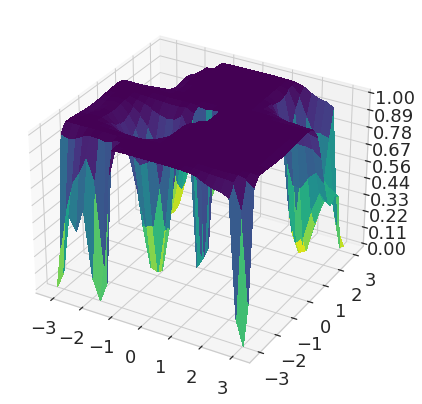}\par
     \caption*{(4,2)}\label{fig:10Q_RN_4_2}
    \hspace*{-0.5cm}
      \includegraphics[scale=0.21]{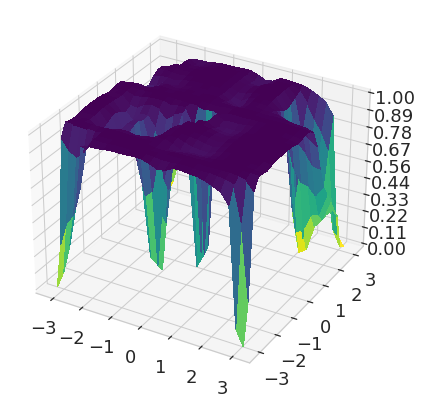}\par
      \caption*{(3,3)}\label{fig:10Q_RN_3_3}
\end{multicols}
\captionsetup{hypcap=false}
\caption{Cost function landscapes of PlainQNet (upper panel) and ResQNet (lower panel) for $10$ Qubits. The parentheses denotes the $D_L$ per QN.}
\label{fig:cf_landscape_PN_RN_Q=10}
\end{figure*}

To further expand our study, we increased the number of qubits to $10$ and performed the same experiments as with quantum layers of $6$ and $8$ qubits. The cost function landscapes were then analyzed for both PlainQNets and ResQNets, as shown in Fig. \ref{fig:cf_landscape_PN_RN_Q=10}. Similar to the case of 8 qubit layers, a substantial portion of the cost function landscape of PlainQNets was found to be flat, indicating the presence of BP and making it unsuitable for optimization. Conversely, the cost function landscape of ResQNets remained more favorable for optimization as it was characterized by multiple paths leading to the global minimum, thus avoiding the occurrence of BP.  

\begin{figure}[h!]
     
        \begin{subfigure}[b]{0.2\textwidth}
         \hspace*{-2.3cm}
        \includegraphics[scale=0.27]{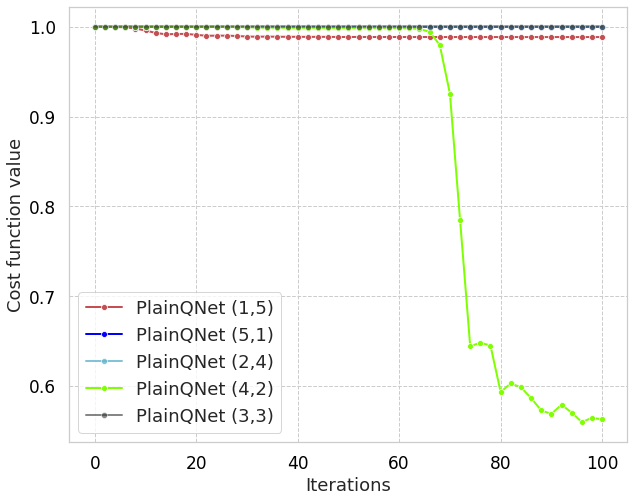}
         \caption{}
         \label{fig:training_PN_10Q}
     \end{subfigure}
       \hfil\quad
   \begin{subfigure}[b]{0.2\textwidth}
         \hspace*{-1.2cm}
         \includegraphics[scale=0.27]{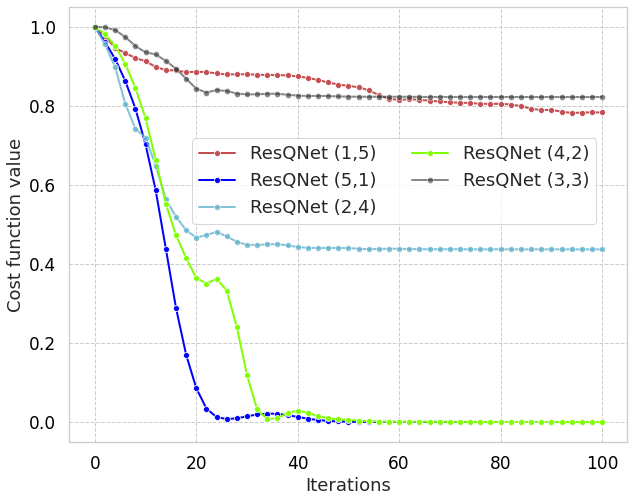}
         \caption{}
         \label{fig:training_RN_10Q}
     \end{subfigure}
     \captionsetup{hypcap=false}
    \caption{Cost vs. iterations of (a) PlainQNets (b) and ResQNets for $10$ qubits. The parantheses denotes the $D_L$ per QN.}  
    \label{fig:training_PN_RN_10Q}
    
\end{figure}

We subsequently trained the 10 qubit quantum layers to address the problem defined in Eq. (\ref{eq:CF_eq}). The results of these experiments are depicted in Fig. \ref{fig:training_PN_RN_10Q}. Our analysis indicates that PlainQNets did not exhibit successful training outcomes for nearly all depth combinations, with the exception of $(4,2)$, which showed considerable performance improvement. When we examined its cost function landscape in Fig. \ref{fig:cf_landscape_PN_RN_Q=10}, we observe that there are one or two narrow regions that contain the solution and may be found by the optimizer. However, these narrow regions are unlikely to be encountered and thus the performance, despite being optimal, is not considered suitable for general optimization problems. Therefore, it can still be concluded that the PlainQNets are severely affected by the problem of BP.  
On the other hand, ResQNets effectively overcame the issue of BP and exhibited successful training outcomes for all depth combinations.
Our observations for 10 qubit quantum layers align with our previous findings for 6 and 8 qubit layers in that ResQNets are more effective when the depth after the residual connection is less. This suggests that a shallower depth of quantum layers after the residual connection in ResQNets is more favorable for optimization and mitigating the impact of BP.

Our results conclusively demonstrate that PlainQNets are heavily impacted by the issue of BP as the number of qubits increases, which significantly hinders their performance and ability to optimize the cost function. 
The previous results have demonstrated the advantage of our proposed ResQNets over PlainQNets in mitigating the phenomenon of BP. Therefore, in the next section, we will conduct experiments solely with ResQNets.

\subsection{ResQNets with wider quantum layers} \label{subsec:RN_wider}
To analyze the scalability of ResQNets for larger quantum circuits, we consider quantum layers with larger number of qubits, i.e., $15$ and $20$. The depth of the quantum layers, $D_L$, is kept constant at $6$. As the cost function landscapes are known to have a direct impact on the training results, as shown in Sect. \ref{sec:RN_shallow}. Consequently, we only present the training results for the 15 and 20-qubit quantum layers.
\subsubsection{15-Qubit Circuit}
We train the 15 qubit quantum layers to optimize the problem defined in Eq. (\ref{eq:CF_eq}). The training results are shown in Fig. \ref{fig:training_RN_15Q}. it can be observed that the ResQNets are effectively trained. Additionally, analogous to the case of shallow width quantum layers, the performance is substantially better when the depth in the first QN (before the residual point) is bigger than the second QN. 

\subsubsection{20-Qubit Circuit}
\begin{figure}[h!]
     
        \begin{subfigure}[b]{0.2\textwidth}
         \hspace*{-2.25cm}
        \includegraphics[scale=0.27]{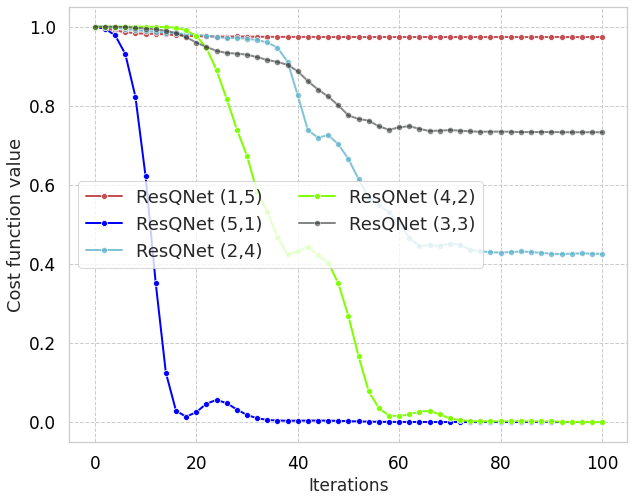}
         \caption{}
         \label{fig:training_RN_15Q}
     \end{subfigure}
       \hfil\quad
   \begin{subfigure}[b]{0.2\textwidth}
         \hspace*{-1.2cm}
         \includegraphics[scale=0.27]{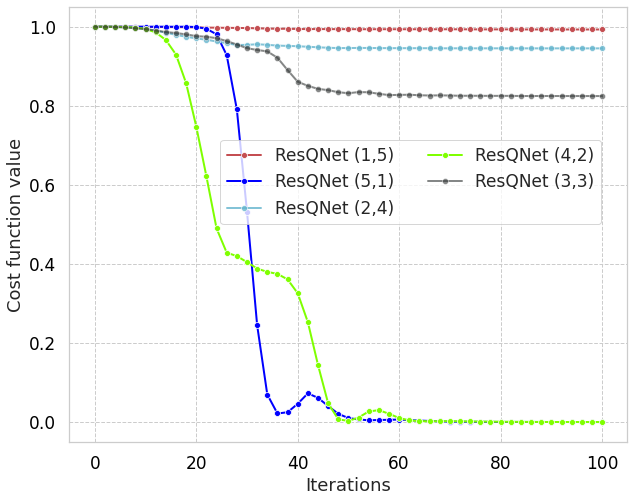}
         \caption{}
         \label{fig:training_RN_20Q}
     \end{subfigure}
     \captionsetup{hypcap=false}
    \caption{Cost vs. iterations of ResQNets for (a) $15$ qubits and (b) $20$ qubits. The parentheses denote the $D_L$ per QN. }  
    \label{fig:training_RN_10Q_20Q}
    
\end{figure}

We now train the ResQNets for 20-qubit layers for the problem defined in Eq. (\ref{eq:CF_eq}), with a total layer depth of $D_L=6$. It can be observed that even with 20 qubit layers, the ResQNets are effectively trained, as shown in Fig. \ref{fig:training_RN_20Q}. Furthermore, similar to the previously shown results, the ResQNets for 20-qubit layers also perform significantly better when the depth after the residual point (second QN) is lesser than the depth before the residual point (first QN).

From the results in Fig. \ref{fig:training_RN_10Q_20Q}, it is evident that the ResQNets are capable of working with wider quantum layers. 
The results demonstrate that analogous to the case of shallow-width quantum layers, the training performance is better with the optimal results being achieved for a larger depth in the first QN and a smaller depth in the second QN.

It should be noted that our experiments are limited by the memory constraints of our local computer and we cannot go beyond 20 qubits. However, based on our findings, we believe that the proposed ResQNets would still train effectively even beyond 20 qubits. 


\subsection{ResQNets with 3-QN}

From the analysis presented in previous sections, it can be observed that the ResQNets consisting of two QNs with a maximum of one residual block can effectively address the problem of BP and significantly improve the training performance of QNNs. In this section, we show that increasing the number of QNs in ResQNets can enhance the performance of ResQNets even further.
%
%
As discussed in Sect. \ref{sec:methodology}, for three QNs we can have multiple configurations of residual blocks. We consider all of these configurations for our experiments with 20-qubit quantum layers and a fixed quantum layer depth of $D_L=6$.
%
The results of the experiments conducted in this section will provide valuable insights into the optimal configuration of residual blocks for ResQNets with three or more QNs.

\begin{figure*}[h!]
\centering
\begin{multicols}{3}
    \includegraphics[scale=0.28]{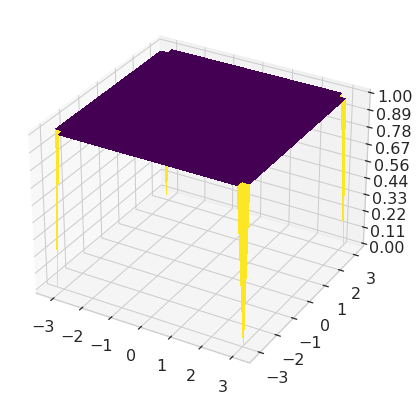}\par
    \caption*{(2,3,1)}\label{fig:3QN_RN_2_3_1}
    
    \includegraphics[scale=0.28]{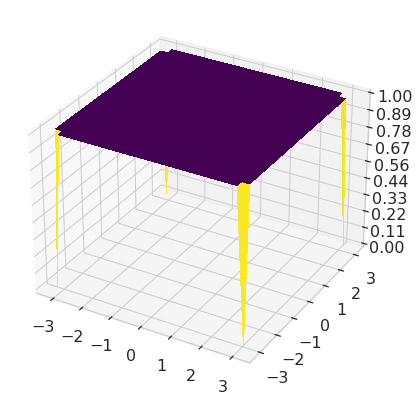}\par
    \caption*{(3,2,1)}\label{fig:3QN_RN_3_2_1}
    
     \includegraphics[scale=0.28]{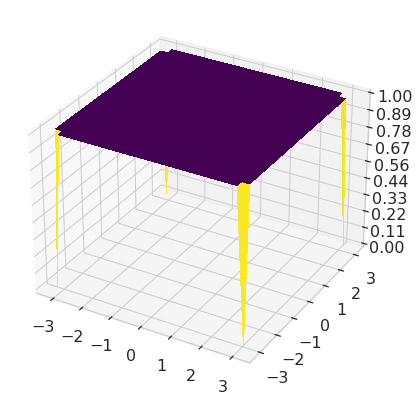}\par
     \caption*{(4,1,1)}\label{fig:3QN_RN_4_1_1}

\end{multicols}

\begin{multicols}{3}
    \includegraphics[scale=0.28]{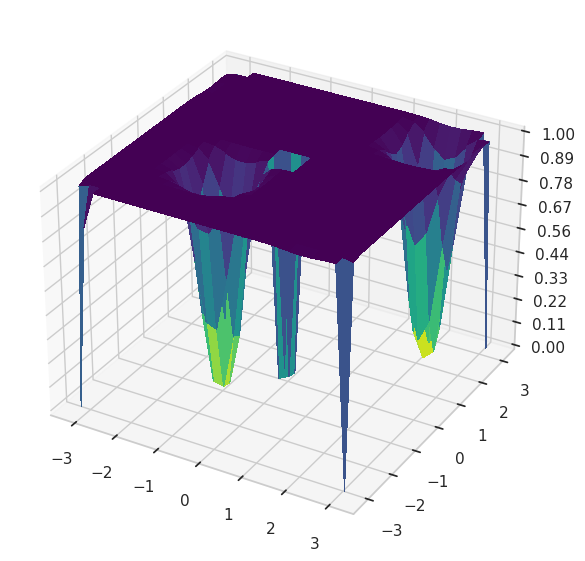}\par
    \caption*{(2 3,1)}\label{fig:3QN_RN_23_1}
    
    \includegraphics[scale=0.28]{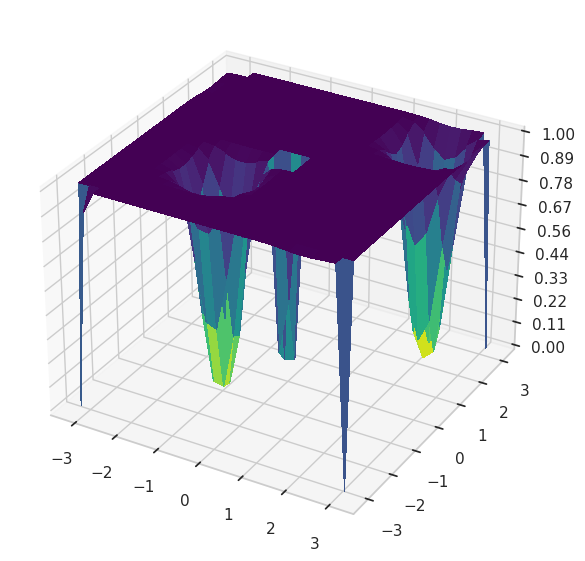}\par
    \caption*{(3 2,1)}\label{fig:3QN_RN_32_1}
    
     \includegraphics[scale=0.28]{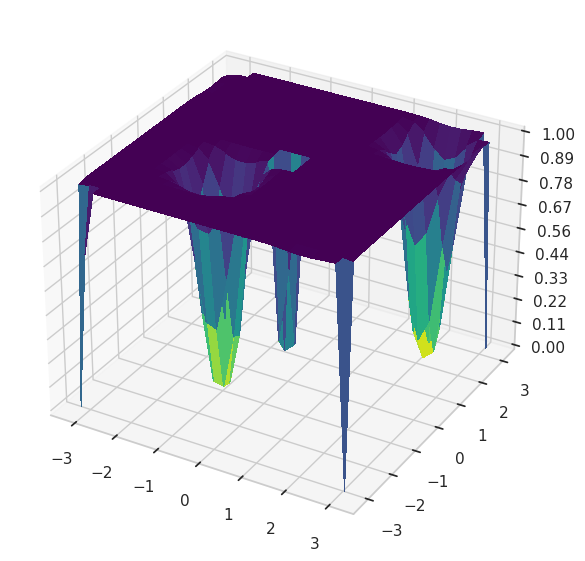}\par
     \caption*{(4 1,1)}\label{fig:3QN_RN_41_1}

\end{multicols}

\begin{multicols}{3}
    \includegraphics[scale=0.28]{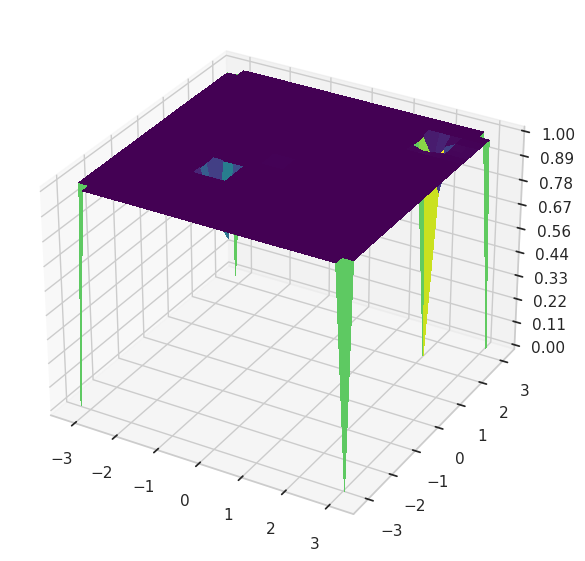}\par
    \caption*{(2,3 1)}\label{fig:3QN_RN_2_31}
    
    \includegraphics[scale=0.28]{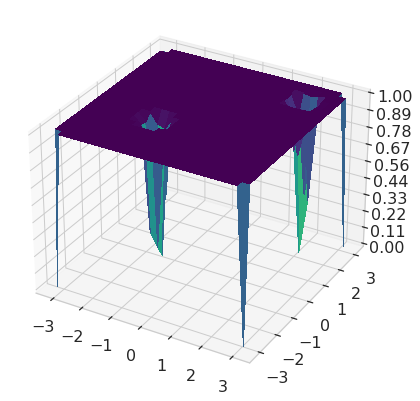}\par
    \caption*{(3,2 1)}\label{fig:3QN_RN_4_11}
    
     \includegraphics[scale=0.28]{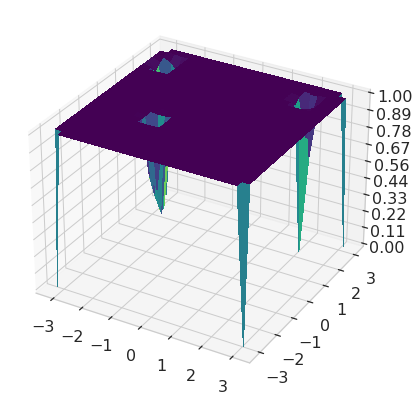}\par
     \caption*{(4,1 1)}\label{fig:3QN_RN_3_21}

\end{multicols}
\captionsetup{hypcap=false}
\caption{Cost function landscapes of ResQNets for $20$ Qubits and 3-QNs. Residual after every QN (Top panel), Residual after two QNs (middle panel) and residual only after the first QN (bottom panel). The parentheses denote the $D_L$ per QN and the comma denotes the residual point.}
\label{fig:cf_landscape_RN_3QN_20Q}
\end{figure*}

The cost function landscapes of various residual block configurations in ResQNets with three QNs were analyzed, as presented in Fig. \ref{fig:cf_landscape_RN_3QN_20Q}. The results indicate that the optimal placement of residual blocks has a significant impact on the performance of ResQNets.
When the residual block is added after every QN, the cost function landscape quickly flattens irrespective of the depth per QN, suggesting that this configuration leads to equivalent or suboptimal performance compared to PlainQNets, which is not at all suitable for optimization.  

On the other hand, when the residual block is added after two QNs, the cost function landscape shows multiple and wider regions containing the global minimum, which makes this configuration more suitable for optimization. Moreover, this configuration exhibits a consistent cost function landscape regardless of the depth per QN combination, implying that this particular residual block arrangement is more robust to BP and supports a wide range of depths and QN combinations. 

For the case of adding the residual only after the first QN, with two QNs after the residual block, the results show that the cost function landscape is better than the case of adding the residual block after every QN, but not as good as the case where there is a gap of two QNs while adding the residual. 

\begin{figure}[h!] 
   \begin{subfigure}{0.3\textwidth}
       \includegraphics[width=\linewidth]{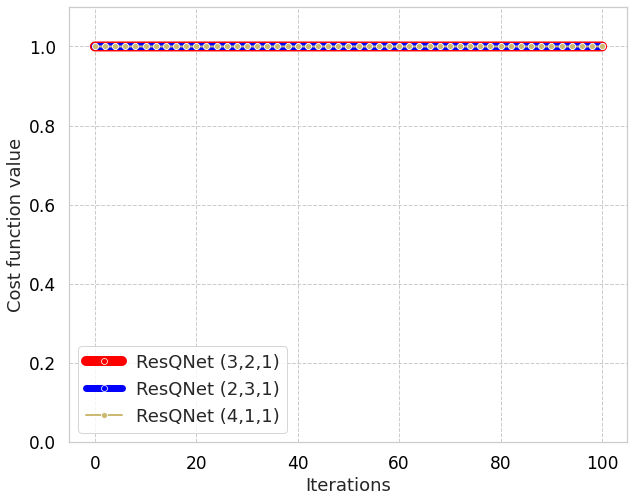}
       \caption{}
       \label{fig:training_3QN_20Q_res_after_ever_QN}
   \end{subfigure}
\hfill
   \begin{subfigure}{0.3\textwidth}
       \includegraphics[width=\linewidth]{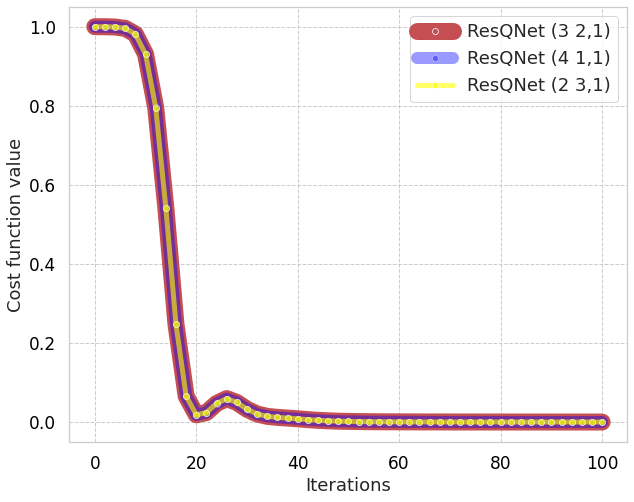}
       \caption{}
       \label{fig:training_3QN_20Q_res_after_two_QN}
   \end{subfigure}
\hfill
   \begin{subfigure}{0.3\textwidth}
       \includegraphics[width=\linewidth]{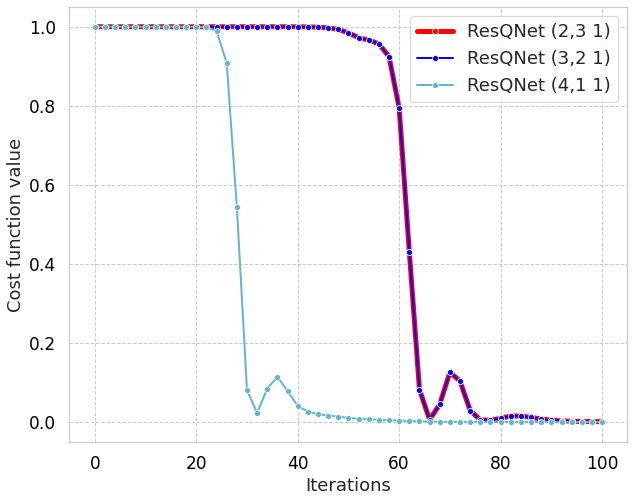}
       \caption{}
       \label{fig:training_3QN_20Q_res_after_first_QN}
   \end{subfigure}
   \captionsetup{hypcap=false}
 \caption{Training results of ResQNets with three QNs with 20 qubit layers. (a) Residual after every QN (b) Residual after two QNs and (c) Residual after the first QN.  The parentheses denote the $D_L$ per QN and the comma denotes the residual point.}  
    \label{fig:training_all_RN_20Q_3_QN}
\end{figure}

We then trained ResQNets with three QNs for all the configurations while varying the depth for each QN combination on the problem defined in Eq. (\ref{eq:CF_eq}). The training results are shown in Fig. \ref{fig:training_all_RN_20Q_3_QN}. These results align with the behavior of the cost function landscape, where the residual block configuration skipping two QNs outperforms other configurations. It can be observed that the residual block configuration after every QN does not train at all, while the residual block configuration after the first QN does converge for all the depth per QN combinations, but with significantly slower convergence compared to the residual block configuration after two QNs.

\subsection{3-QN vs. 2-QN ResQNet}
In this section, we compare the performance of ResQNets with 2 and 3-QNs to demonstrate the impact of increasing the number of QNs. The analysis was conducted for 20 qubit layers considering the best-performing depth combinations for both 2 and 3-QNs.

\begin{figure}[h]

     \centering
     
        \begin{subfigure}[b]{0.2\textwidth}
         \hspace*{-2.3cm}
        \includegraphics[scale=0.27]{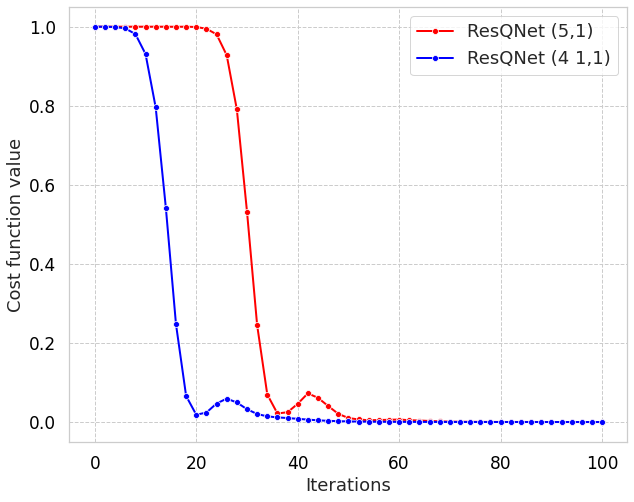}
         \caption{}
         \label{fig:training1_RNs_20Q_2_3QN}
     \end{subfigure}
       \hfil\quad
   \begin{subfigure}[b]{0.2\textwidth}
         \hspace*{-1.2cm}
         \includegraphics[scale=0.27]{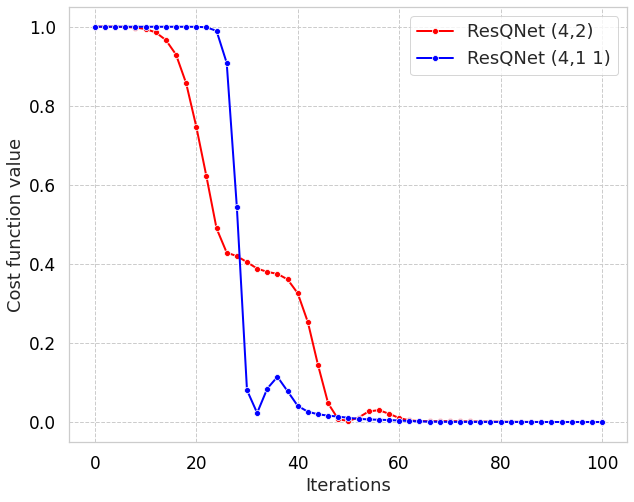}
         \caption{}
         \label{fig:training2_RNs_20Q_2_3QN}
     \end{subfigure}
     \captionsetup{hypcap=false}
    \caption{Training comparison of 2-QN and 3-QN ResQNets for 20 qubit layers. The parentheses denote the $D_L$ per QN and the comma denotes the residual point.}  
    \label{fig:training_RN_20Q_2_3_QN}
    
\end{figure}

For 2-QNs, the results from Fig. \ref{fig:training_RN_20Q} indicate that the depth combinations of (5,1) and (4,2) performed better than other depth combinations. On the other hand, for three QNs, the results from Fig. \ref{fig:training_3QN_20Q_res_after_two_QN} and \ref{fig:training_3QN_20Q_res_after_first_QN} show that the depth combinations of $(4\hspace{0.1cm}1,1)$ and $(4, 1\hspace{0.1cm}1)$ outperformed other depth combinations.
A closer examination of the best-performing depth combinations reveals that the $D_L$ before and after the residual block for the depth per QN combination of $(5,1)$ in 2-QN ResQNet is equivalent to depth per QN combination of $(4\hspace{0.1cm} 1,1)$ for 3-QN ResQNet. Similarly, the combination $(4,2)$ in the 2-QN ResQNet is equivalent to $(4,1\hspace{0.1cm} 1)$ in the 3-QN ResQNet. Despite these similarities, as demonstrated in Fig. \ref{fig:training_RN_20Q_2_3_QN}, the ResQNets with 3-QNs exhibit superior performance, as they converge to the optimal solution more efficiently compared to the ResQNets with 2-QNs.

\subsection{Real Quantum Device}
\begin{figure}[h!]
     
        \begin{subfigure}[b]{0.2\textwidth}
        \hspace*{-2.3cm}
        \includegraphics[scale=0.27]{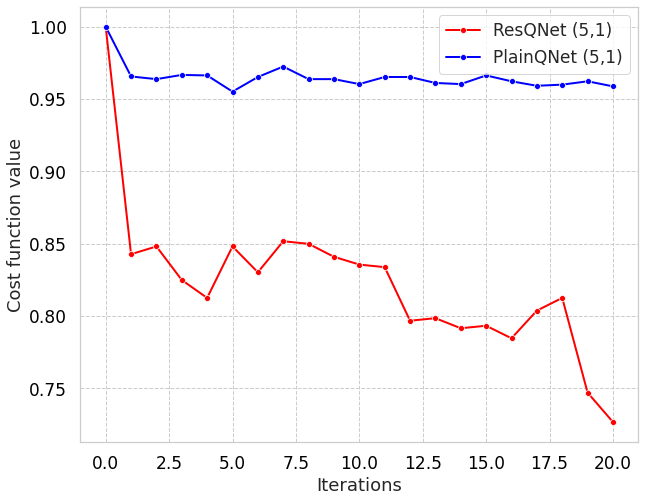}
         \caption{}
         \label{fig:training_5q_PN_RN_real_device}
     \end{subfigure}
       \hfil\quad
   \begin{subfigure}[b]{0.2\textwidth}
         \hspace*{-1.25cm}
         \includegraphics[scale=0.27]{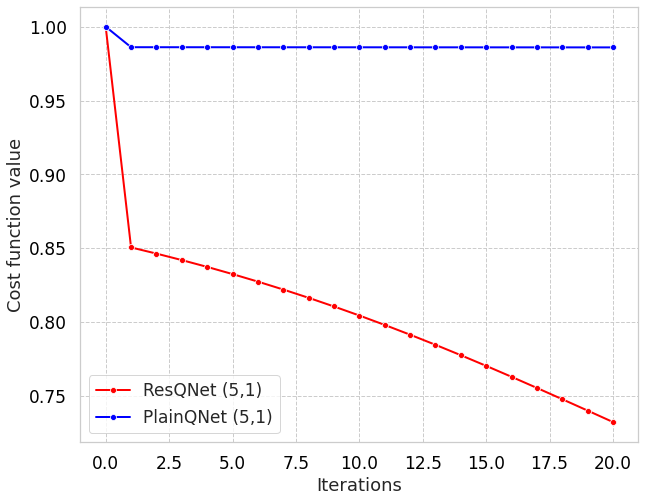}
         \caption{}
         \label{fig:training_5q_PN_RN_sim}
     \end{subfigure}
     \captionsetup{hypcap=false}
    \caption{Training comparison ResQNets and PlainQNets on (a) real quantum device and (b) simulator. The values in parentheses denote the depth per QN.}  
    \label{fig:training_RN_PN_realdevice}
    
\end{figure}

The results presented so far were obtained by running ResQNets and PlainQNets on a simulation platform.
In this section, we carry out some experiments on real quantum devices. In particular, we trained both ResQNets and PlainQNets with 2-QNs on a 5-qubit quantum layer with 20 epochs using an IBM's quantum device, namely $ibmq\_lima$. The quantum layers depth was fixed to $D_L=6$ with $D_L=5$ in the first QN, and $D_L=1$ in the second QN. This depth combination was chosen considering all the results discussed previously.  We note that due to the limited number of publicly available quantum devices, the queue times for executing the jobs are considerably long. 
Therefore, to minimize the training time, we chose to reduce the number of epochs for real-device training. We trained both PlainQNets and ResQNets for only $20$ epochs on real devices instead of $100$ epochs as in the case of simulation. The training results are illustrated in Fig. \ref{fig:training_RN_PN_realdevice}.

The results presented in Fig. \ref{fig:training_5q_PN_RN_real_device} reveal that ResQNets have been trained successfully on a real device, whereas PlainQNets have not been trained on a real device. The same trend is observed when both networks are executed on the simulator, as depicted in Fig. \ref{fig:training_5q_PN_RN_sim}.
However, when both PlainQNets and ResQNets are trained on a real device, a slight fluctuation is observed while approaching the optimal solution due to hardware noise, as compared to the simulation results. 
Despite the presence of noise, the rate of decrease in the loss value for ResQNets is almost identical for both simulation and real experiments. 
According to \cite{Wang:2021}, hardware noise can potentially cause BP. However, our results demonstrate that our proposed ResQNets are somewhat resilient against hardware noise, as they achieve similar performance to that of the simulator (though with some fluctuations).



\section{Conclusion}\label{sec:conclusion}
The problem of barren plateaus (BP) in quantum neural networks (QNNs) is a critical hurdle on the road to the practical realization of QNNs. 
There have been several attempts to resolve this issue, but the impact of BP can still vary greatly depending on the application and the architecture of quantum layers. Thus, it is essential to have multiple solutions for BP to cover a wide range of problems.

In this paper, we propose residual quantum neural networks (ResQNets) to address the issue of BP in QNNs. Our approach is inspired by classical residual neural networks (ResNets), which were introduced to overcome the vanishing gradients problem in classical neural networks.

In traditional QNNs, a single parameterized quantum circuit (PQC) with arbitrary depth is included within a single quantum node (QN). To create ResQNets, we split the conventional QNN architecture into multiple QNs, each containing its own PQC with varying depths. Splitting the QNNs allows us to introduce the residual connections between the QNs, forming our proposed ResQNets.
In simple QNNs without residual connections (referred to as PlainQNets), the output from the previous QN serves as the input to the next. 
On the other hand, in ResQNets, one or multiple QNs can serve as residual blocks, with the output from a previous residual block being added to its input before it is passed on to the next QN.

In our study, we first demonstrate the efficacy of the proposed splitting of the conventional QNN architecture into multiple QNs (PlainQNets) by comparing their performance to that of conventional QNNs (simple PlainQNets). The comparison results indicated that the PlainQNets have better or equivalent performance to that of conventional QNNs. Subsequently, we compare the performance of PlainQNets with that of our proposed ResQNets through several training experiments. Our analysis of the cost function landscapes for quantum layers of increasing qubits shows that incorporating residual connections results in improved training performance.

Based on our findings, we conclude that the proposed ResQNets provide a promising solution for overcoming the problem of BP in QNNs and offer a potential direction for further research in the field of quantum machine learning.

\begin{backmatter}

\section*{Acknowledgements}
The authors thank Qatar National Library for supporting the open access publication of this article.

\section*{Funding}
Not Applicable.

\section*{Abbreviations}
QNN, Quantum Neural Network; BP, Barren Plateaus; QN, Quantum Node; ResNet, Classical Residual Neural Network; ResQNet, Quantum Residual Neural Network; PQC, Parameterized Quantum Circuit.

\section*{Availability of data and materials}
The code that was used to generate the numerical results in this work can be provided on request.

\section*{Ethics approval and consent to participate}
Not Applicable.

\section*{Competing interests}
The authors declare that they have no competing interests.

\section*{Consent for publication}
Both authors have approved the publication. The research in this work did not involve any human, animal or other participants.

\section*{Authors' contributions}
Both the authors contributed to the work described in this paper. M.K conceived the idea and performed the numerical experiments, and prepared the manuscript. S.A.K reviewed the manuscript and approved the submission. 



\bibliographystyle{bmc-mathphys} 
\bibliography{bmc_article}      


\end{backmatter}
\end{document}